\documentclass[a4paper, notoc]{JHEP3}
\usepackage[dvips]{graphicx}
\usepackage{amsfonts}
\usepackage{amsmath}
\usepackage{amssymb}
\usepackage{epsfig}
\usepackage{cite}
\usepackage{multicol}
\usepackage{axodraw}
\usepackage{subfigure}
\DeclareGraphicsExtensions{.eps}

\def\eq{\begin{equation}}
\def\qe{\end{equation}}
\def\eqa{\begin{eqnarray}}
\def\qea{\end{eqnarray}}
\newcommand{\newc}{\newcommand}
\newc{\softsusy}{\texttt{SOFTSUSY}}
\newc{\looptools}{\texttt{LoopTools}}
\newc{\isajet}{\texttt{ISAJET7.64}}

\newc{\RPVSUGRA}{RPV mSUGRA}

\newc{\MO}{m_0}
\newc{\Mhalf}{M_{1/2}}
\newc{\AO}{A_0}
\newc{\tanb}{\textrm{tan}\beta}
\newc{\sgnmu}{\textrm{sgn}{\mu}}

\newc{\bsmumu}{BR(B_s\to \mu^+\mu^-)}
\newc{\bsgamma}{BR(b\to s \gamma)}
\newc{\MX}{M_X}
\newc{\MZ}{M_Z}

\newc{\lag}{\mathcal{L}}
\newc{\mM}{\mathcal{M}}
\newc{\mnu}{\m^{\nu}}

\newc{\ssup}{\tilde{\bar{U}}}
\newc{\ssdown}{\tilde{\bar{D}}}
\newc{\ssstrange}{\tilde{s}}
\newc{\sscharm}{\tilde{c}}
\newc{\sstop}{\tilde{t}}
\newc{\ssbottom}{\tilde{b}}

\newc{\sse}{\tilde{\bar{E}}}
\newc{\ssmu}{\tilde{\mu}}
\newc{\sstau}{\tilde{\tau}}
\newc{\ssnue}{\tilde{\nu_{e}}}
\newc{\ssnumu}{\tilde{\nu_{\mu}}}
\newc{\ssnutau}{\tilde{\nu_{\tau}}}
\newc{\ssbnue}{\bar{\tilde{\nu_{e}}}}
\newc{\ssbnumu}{\bar{\tilde{\nu_{\mu}}}}
\newc{\ssbnutau}{\bar{\tilde{\nu_{\tau}}}}
\newc{\ssnu}{\tilde{\nu}}

\newc{\lam}{\lambda}
\newc{\sslep}{\tilde{L}}
\newc{\ssq}{\tilde{Q}}
\newc{\ssh}{\tilde{h}}

\newc{\neut}{\tilde{\chi}^0}
\newc{\charge}{\tilde{\chi}}
\newc{\glu}{\tilde{g}}

\newc{\higgs}{h^0}
\newc{\Higgs}{H^0}
\newc{\Azero}{A^0}

\newc{\nue}{\nu_e}
\newc{\numu}{\nu_{\mu}}
\newc{\nutau}{\nu_{\tau}}
\newc{\bnue}{\bar{\nu_e}}
\newc{\bnumu}{\bar{\nu_{\mu}}}
\newc{\bnutau}{\bar{\nu_{\tau}}}

\newc{\intd}{\int\frac{d^Dq}{(2\pi)^D}}
\newc{\sig}{\sigma}
\newc{\sigbar}{\bar{\sigma}}
\newc{\gmunu}{g_{\mu\nu}}
\newc{\qmu}{q_{\mu}}
\newc{\qnu}{q_{\nu}}
\newc{\kmu}{k_{\mu}}
\newc{\knu}{k_{\nu}}
\newc{\Chi}{\xi}

\newcommand{\be}{\begin{equation}}
\newcommand{\ee}{\end{equation}}
\newcommand{\bea}{\begin{eqnarray}}
\newcommand{\eea}{\end{eqnarray}}

\newcommand{\lsim}{\lesssim}

\newcommand{\beqa}{\begin{eqnarray}}
\newcommand{\eeqa}{\end{eqnarray}}

\def\bib{\bibitem}

 \newlength{\wth}
 \title{Lepton number violating mSUGRA and neutrino masses}
\author{B.~C. Allanach, C.~H. Kom \\ 
  DAMTP, Centre for Mathematical Sciences,
  Wilberforce Road, Cambridge, CB3 0WA, United Kingdom\\
 E-mail:
   \email{b.c.allanach@damtp.cam.ac.uk}, \email{c.kom@damtp.cam.ac.uk}
}

\abstract{
We perform a quantitative study of neutrino phenomenology in the framework
of minimal supergravity (mSUGRA) with grand unified theory (GUT)-scale
 tri-linear lepton number violation.  
We show that only two non-zero GUT scale lepton number
violating parameters and three charged lepton mixing angles are sufficient to account for current neutrino
oscillation data.  This allows collider studies to be performed in a
manageable parameter space.  We discuss some phenomenological consequences of
the models, including tuning issues.  
}

\preprint{DAMTP-2007-106}


\begin{document}

\tableofcontents

\section{Introduction}

In general, the minimal supersymmetric standard model (MSSM) contains
baryon and lepton number violating (LNV) operators in the
superpotential~\cite{ssm-superpot}.  This 
generally leads to fast proton decay beyond experimental limits, unless an
additional symmetry is imposed 
on the theory.  The most widely studied symmetry is R-parity, which
forbids all problematic renormalisable operators and makes
the lightest supersymmetric particle (LSP) stable, providing a potential dark
matter candidate.  However, there are still
dimension five operators which may be dangerous for proton
decay~\cite{proton-decay}. If one instead imposes an anomaly-free proton
hexality discrete abelian symmetry $P_6$~\cite{Dreiner:2005rd}, such dimension
five operators are forbidden. 
Another alternative to ensure slow enough proton decay is so-called baryon
triality, which is an anomaly-free $\mathcal{Z}_3$ symmetry
\cite{Ibanez:1991hv} and forbids baryon
number violating terms past dimension five.  
The model allows a subset of all available
MSSM R-parity violating operators~\cite{0406039}, namely the LNV operators.
LNV operators in turn
lead to the generation of neutrino masses.  Thus, the LNV MSSM provides an
alternative neutrino mass generation mechanism to the see-saw mechanism 
\cite{Minkowski, Ramond}, in
which gauge singlet right-handed neutrino superfields are added to the
$P_6-$conserving MSSM. 
Although the LSP becomes unstable in such schemes, it could still form
a dark matter candidate if it is the gravitino, since the decays are slow on
cosmological time scales~\cite{Buchmuller:2007ui}.  
Here, we shall neglect the dark matter 
relic density, since it could instead originate from a hidden sector.
The aim
of this paper is to find and investigate simple GUT-scale LNV MSSM models
that can give rise to the observed neutrino mass and mixing pattern.  

The chiral superfields of the MSSM are charged under the standard model (SM) gauge
group $G_{SM}=SU(3)_c \otimes SU(2)_L \otimes U(1)_Y$ as 
\eq
\begin{array}{rclrclrcl}
Q&:&(3,2,\frac{1}{6}),\qquad & \bar{U}&:&(\bar{3},1,-\frac{2}{3}),\qquad & \bar{D}&:&(\bar{3},1,\frac{1}{3}),\qquad \\
L&:&(1,2,-\frac{1}{2}), & H_d&:&(1,2,-\frac{1}{2}), &&& \\
\bar{E}&:&(1,1,1), & H_u&:&(1,2,\frac{1}{2}). &&&
\end{array}
\qe
The full renormalisable LNV MSSM superpotential is given by
\eqa
\mathcal{W}&=&\mathcal{W}_{RPC}+\mathcal{W}_{LNV},
\qea
where $\mathcal{W}_{RPC}$ is the superpotential in the R-parity conserving (RPC) case, and $\mathcal{W}_{LNV}$ is the LNV superpotential given by
\eqa \label{eq:superpotentials}
\mathcal{W}_{RPC}&=&(Y_E)_{ij}H_dL_i\bar{E_j} +(Y_D)_{ij}H_dQ_i\bar{D_j} +(Y_U)_{ij}Q_iH_u\bar{U_j} -\mu H_dH_u, \nonumber \\
\mathcal{W}_{LNV}&=&\frac{1}{2}\lam_{ijk}L_iL_j\bar{E_k}
+\lam'_{ijk}L_iQ_j\bar{D_k} -\mu_iL_iH_u, \label{superpot}
\qea
where we have suppressed gauge indices and $\{ i,j,k \} \in \{1,2,3\}$ are
family indices.
$(Y_E)$, $(Y_D)$ and $(Y_U)$ are $3 \times 3$ matrices of dimensionless Yukawa
couplings; 
$\lambda_{ijk}$, $\lambda'_{ijk}$  are the dimensionless
tri-linear LNV couplings, and $\mu_i$ are dimensionful bi-linear LNV
parameters.   
The soft supersymmetry (SUSY) breaking Lagrangian with LNV is given by
\eqa
-\mathcal{L}_{soft}&=&\mathcal{L}^{mass}_{RPC} +\mathcal{L}_{RPC}^{int}
+\mathcal{L}_{LNV}, 
\qea
where
\eqa\label{eq:scalar potential}
\mathcal{L}^{mass}_{RPC}&=&\frac{1}{2}M_1\tilde{B}\tilde{B} +\frac{1}{2}M_2\tilde{W}\tilde{W} +\frac{1}{2}M_3\tilde{g}\tilde{g} +h.c. \nonumber \\
&&+\ssq^{\dagger}(m^2_{\ssq})\ssq  +\ssup^{\dagger}(m^2_{\ssup})\ssup  +\ssdown^{\dagger}(m^2_{\ssdown})\ssdown  +\sslep^{\dagger}(m^2_{\sslep})\sslep  +\sse^{\dagger}(m^2_{\sse})\sse \nonumber \\
&&+m^2_{h_u}h_u^{\dagger}h_u +m^2_{h_d}h_d^{\dagger}h_d, \nonumber \\
\mathcal{L}_{RPC}^{int}&=&(h_E)_{ij}h_d\sslep_i\sse_j +(h_D)_{ij}h_d\ssq_i\ssdown_j +(h_U)_{ij}\ssq_ih_u\ssup_j -\tilde{B}h_dh_u + h.c.,\nonumber \\
\mathcal{L}_{LNV}&=&\frac{1}{2}h_{ijk}\sslep_i\sslep_j\sse_k +h'_{ijk}\sslep_i\ssq_j\ssdown_k -\tilde{D}_i\sslep_ih_u \nonumber \\
&&+h_d^{\dagger}m^2_{H_dL_i}\tilde{L}_i + h.c. \label{softpot}
\qea
where tilde denotes a super-partner of the more familiar Standard Model field.  $M_1$, $M_2$ and $M_3$ are the masses of the super-partner of the SM gauge bosons, the $m^2$'s are the scalar mass parameters of the sparticle and/or higgs fields, and the $h$'s, $\tilde{B}$ and $\tilde{D}_i$ are trilinear SUSY breaking parameters that correspond to the dimensionless supersymmetric parameters displayed in Eq.~(\ref{eq:superpotentials}).  Tree level neutrino masses originating from such a theory
were derived in Refs.~\cite{9508271, 9408224}. 
Some of the weak-scale parameters in Eqs.~(\ref{superpot},\ref{softpot}) were
bounded by neutrino oscillation data in Refs.~\cite{0007041,0208009}.
The bi-linear couplings have been employed to fit solar and atmospheric
neutrino oscillations at the weak
scale~\cite{Romao:1999up,Hirsch:2000ef,0302021}.  
Work on bi-linear couplings in the MSSM can also be found in Refs.\cite{9901254, 0111332, 9606388}.
Early attempts to calculate tri-linear coupling contributions to the neutrino masses
 can be found in \cite{9702421, 9810536, 0311310, 0005080, 9903435}.  A basis independent calculation was performed in \cite{Davidson:2000ne}.
The complete set of 1-loop corrections to neutrino masses and mixings were investigated in detail in \cite{0603225}.
There, collections of weak-scale couplings that could give 
rise to the tri-bi maximal mixing pattern \cite{Harrison} were suggested. Such 
a pattern provides a successful fit to oscillation constraints on lepton mixing. 

The number of parameters in Eq.~(\ref{softpot}) may be reduced by assumptions
about the structure and origin of the supersymmetry breaking terms. We take the
concrete example of the mSUGRA assumption\footnote{Also termed the constrained
  MSSM.}, where at a high energy GUT scale
$\MX \sim 10^{16}$ GeV the scalar masses are set to be diagonal and equal
to $m_0$, all soft trilinear couplings are equal to their analogous
superpotential parameter multiplied by a common coupling $A_0$ and all gaugino
masses are set to be equal to $M_{1/2}$. Constraints upon such a model
including LNV effects have been studied in
Refs.~\cite{9906209,0309196}. In the latter, a cosmological bound $\sum_i m_{\nu_i}<0.7$ eV was
placed upon the model but no attempt was made to fit the neutrino
properties deduced from oscillation measurements.  In Ref.~\cite{0507311}, three relatively large $\lam'_{ijk}$ defined in a weak basis at the GUT scale were used to generate $\kappa_i$ and $\lam'_{i33}$ at the weak scale.  These couplings are required to satisfy upper bounds obtained from neutrino oscillation data.  The input parameters are not directly related to neutrino phenomenology, but lead to interesting collider signals.  However, no detailed fitting of the neutrino oscillation data was attempted.  
Bi-linear contributions to the neutrino masses in LNV mSUGRA were considered
in Ref.~\cite{9511288}. Solutions with 3 non-zero GUT scale LNV parameters $\mu_i$ were found.  
Also, a preference for the vacuum oscillation solution \cite{VO}
to the solar neutrino problem was stated. This has
since been ruled out by KamLAND data \cite{KamLAND}.  
An attempt to fit the neutrino masses in LNV mSUGRA using $\lam_{i33}$ and $\lam'_{i33}$
 in the charged lepton mass basis was made in Ref.~\cite{9807327}
 using the renormalization group equations (RGEs) of Ref.~\cite{9602381}.  
Only the atmospheric mixing angle and 
the neutrino mass squared differences were included in the fit and a 
preference towards the small mixing angle MSW \cite{SMA} 
solution for the solar neutrinos was found.  The latter has also been excluded
by KamLAND, which favours the large angle MSW region of parameter space.

In the present paper, we shall follow an alternative approach. At the GUT
scale and in the weak
interaction basis where all soft terms are diagonal in flavour space, we shall
assume zero bi-linear LNV operators. 
In this basis, we shall introduce a small
number of non-zero tri-linear LNV operators, following
Ref.~\cite{0309196}. Further non-zero bi- and 
tri-linear couplings are generated on renormalisation to the weak scale. One
may hope that a single GUT-scale tri-linear operator could then be enough to
generate neutrino masses.  
We shall find that a single non-zero trilinear coupling is insufficient
 to reproduce the neutrino mass pattern observed.  A further operator shall
 be necessary.  Although we assume only two non-zero LNV couplings, we think
 of this only as a limiting case, where any other LNV parameters contribute negligibly to the neutrino masses and hence are neglected.  
Such flavour structure may arise from some
 fundamental flavour physics models which yield a small number of dominant
 LNV couplings in the weak interaction basis, mirroring the case of the
 Standard Model Yukawa couplings.

The paper proceeds as follows:
section~\ref{sec:analytics} explains the difficulty in obtaining the observed
neutrino mass hierarchy with only one R-parity violating (RPV)\footnote{We
  will use RPV and LNV interchangeably in this paper.} 
 parameter at $\MX$ in any flavour
basis, and suggests how the next to minimal case with two RPV parameters
(defined in a mixed charged lepton basis) may get around the problems.
We then describe a numerical procedure to find
the best fit value for the 2 RPV parameters, as well as the 3 mixing angles which define the mixed charged lepton basis
 in section~\ref{sec:numerics}.  The
results are presented in section~\ref{sec:results}, where we briefly speculate about
 possible collider signatures.
  We compare our paper with selected work in the literature, and discuss
 issues related to parameter tuning before concluding in section~\ref{sec:conclusion}.
  We set our notation in Appendix~\ref{Appendix: mass matrices},
and the general expressions for the 1-loop neutrino-neutralino mass
corrections in Appendix~\ref{Appendix: self energies}.  Appendix~\ref{CPECPO}
describes the mass insertion approximation for CP even and CP odd neutral
 scalar 
contributions (required for numerical stability), and finally in Appendix~\ref{llgamma}
 a full calculation of the branching ratio of $l^I\to l^J \gamma$
is presented, which will be used to constrain the LNV parameters.

\section{Neutrino masses and mixings}\label{sec:analytics}
In a recent global 3 neutrino fit to all oscillation data \cite{0704.1800}, the following ranges of parameters at $1 \sigma$ were found
\begin{eqnarray} \label{eq:neutrino_data}
\Delta m^2_{21} &=&  7.9^{+0.27}_{-0.28}\times10^{-5}\textrm{eV}^2, \qquad
|\Delta m^2_{31}|  =  2.6 \pm 0.2 \times10^{-3}\textrm{eV}^2, \\
\textrm{sin}^2\theta_{12}  &=&  0.31 \pm 0.02,\qquad
\textrm{sin}^2\theta_{23}  = 0.47^{+0.08}_{-0.07},\qquad
\textrm{sin}^2\theta_{13}  =  0^{+0.008}_{-0.0}. \nonumber
\end{eqnarray}
Here $\Delta m^2_{21}$ and $\Delta m^2_{31}$ denote the mass squared differences between the three neutrinos responsible
 for solar and atmospheric oscillations repectively.  The $\textrm{sin}^2\theta$'s represent the rotation angles characterizing the
 PMNS matrix \cite{PMNS}, the lepton counterpart to the CKM matrix, in the standard parameterization \cite{PDG}.
First, we shall discuss 
some issues concerning mixing matrices and effective neutrino and charged lepton mass matrices in the presence of lepton number violation.

Since we work in the weak interaction basis, it is instructive to understand how
diagonalization of the charged lepton Yukawa matrix behaves, particularly under renormalization.
 In the R-parity conserving limit and in the absence of the
neutrino Yukawa matrix, one can simply rotate the leptonic superfields
back to a basis where all the
leptonic terms are flavour diagonal at $\MX$.  The rotation matrices $Z_{lL}$ and $Z_{lR}$
are defined by
\eqa\label{eq:biunitary rotation}
Z_{lL}^{\dagger}Y_E Z_{lR} &=& \hat{Y}_E,
\qea
where $\hat{Y}_E$ is diagonal.  Since mSUGRA soft terms are diagonal in
flavour space, 
there is no intrinsic flavour violation in the RPC limit.  It is therefore
 always possible to diagonalise the
leptonic Yukawa couplings and the slepton mass matrix by the same rotations even after
renormalization to lower scales.  The rotation is also renormalization scale
independent.  

To show this we examine
 the renormalization group (RG) equations of $Y_E$~\cite{Martin:1993zk},
 keeping only terms proportional to Yukawa matrices, as the gauge coupling
 contributions are flavour blind and so cannot change the lepton mixing.  At
 1-loop, we have 
\eqa
\frac{d}{dt}Y_E &=& \frac{1}{(4\pi)^2}Y_E\{\textrm{Tr}(3Y_DY_D^{\dagger}+Y_EY_E^{\dagger})+3Y_E^{\dagger}Y_E + \textrm{g.c.} \} \nonumber \\
&\equiv& aY_E\Big\{ Y_E^{\dagger} Y_E +
 \hat{\beta}\Big(\textrm{Tr}(Y_DY_D^{\dagger}),\textrm{Tr}(Y_EY_E^{\dagger}),g_1,g_2\Big)\Big\}, \label{rgeeq1}
\qea
where g.c.\ stands for gauge contributions and $t=\ln \mu$ is the natural
logarithm of the ${\overline{DR}}$ renormalisation scale. The function
 $\hat{\beta}\Big(\textrm{Tr}(Y_DY_D^{\dagger}),\textrm{Tr}(Y_EY_E^{\dagger}),g_1,g_2\Big)$
 depends on the trace of the quark and lepton Yukawa matrices as well as the
 gauge couplings, and so is proportional to the unit matrix and do not change
 lepton flavours as a result.  

Writing $Y_E(t) \equiv Y_t$, $\hat \beta(t) \equiv {\hat \beta}_t$ and
$Y_E(0)=Y_0$, we obtain 
\eqa
Y_t &=& Y_0 + a\int_0^t dt_1 Y_{t_1} (Y^{\dagger}_{t_1}Y_{t_1}
+ {\hat \beta}_{t_1}). \label{expandsol}
\qea
We may then plug Eq.~(\ref{expandsol}) back into itself in order to obtain an
expansion
\eqa
Y_t
&=& Y_0 + a\int_0^t dt_1
\left[Y_0+a\int_0^{t_1}dt_2Y_{t_2}(Y^{\dagger}_{t_2}Y_{t_2}
+ {\hat \beta}_{t_2})
\right]
 \label{rgeeq2} \\
&&
\left\{
\left[
Y_0+a\int_0^{t_1}dt_2Y_{t_2} (Y^{\dagger}_{t_2}Y_{t_2} + {\hat \beta}_{t_2})
\right]^{\dagger}  
\left[Y_0+a\int_0^{t_1}dt_2
Y_{t_2}(Y^{\dagger}_{t_2}Y_{t_2} + {\hat \beta}_{t_2})\right] + {\hat
  \beta}_{t_1} \right\}. \nonumber
\qea
Eq.~(\ref{expandsol}) may be substituted back into Eq.~(\ref{rgeeq2}) an arbitrary
number of times, but it is clear that all terms are products of 
terms which are either of the form $\int d t_n Y_{t_n}  (Y_{t_n}^\dagger
Y_{t_n}) D(t_n)$ or $\int d t_n Y_{t_n} D'(t_n)$ where $D(t_n)$ and $D'(t_n)$
are some functions of $t_n$ which are diagonal in flavour space.
We may now calculate $Y_t$ to arbitrary order in $a$, for example
\eqa
Y_t &=& Y_0 +a\left[ AY_0  + t Y_0Y_0^{\dagger}Y_0\right] +
a^2 \left[
\frac{3t^2}{2}Y_0Y_0^{\dagger}Y_0Y_0^{\dagger}Y_0 + 
B Y_0 Y_0^\dagger Y_0 + C Y_0 \right] +
\mathcal{O}(a^3),
\qea
where $A=\int^t_0 dt_1 {\hat \beta}_{t_1}$, $B=3\int^t_0 dt_1 \int^{t_1}_0 dt_2
{\hat \beta}_{t_2} + \int^t_0 dt_1 t_1 {\hat \beta}_{t_1}$ and 
$C=\int^t_0 dt_1 {\hat \beta}_{t_1} \int^{t_1}_0 dt_2 {\hat \beta}_{t_2}$.
It is clear that higher order corrections always have the form
$Y_0(Y_0^{\dagger}Y_0)^n$ plus some diagonal piece, and so both $Y_t$ and
$Y_0$ can be diagonalized by the same mixing matrices $Z_{lL(R)}$. 
This procedure can be generalised to higher loops, but we
decline to do so here.
 Intuitively, in a non-diagonal $Y_E$ basis,
  the mass eigenstates are defined
 by the vectors in the lepton mixing matrices $Z_{lL}$ and $Z_{lR}$, and after the rotation
  the new flavour basis coincides with the mass basis.  Because there
 is no intrinsic lepton flavour changing physics in the Lagrangian, these mass
 eigenstates cannot mix with each other under renormalization,
  except through a trace which then affects all three flavours in the same way.  The mixing
 matrices hence remain the same. 

This means that even though we work in a non-diagonal lepton flavour basis,
lepton flavour changing processes only arise from the LNV couplings and are
suppressed by even powers of them since the LNV operators violate lepton
number by one unit and the Majorana neutrino masses generated violate it by
two.  Another source of leptonic flavour changing neutral current (FCNC)
 would arise when one allows 
for non-universal SUSY breaking boundary conditions, which is common in unified model
building.  This is a separate issue and has been discussed extensively in the
literature, see e.g. \cite{9501334}. 

The LNV operators also make the rotation matrices $Z_{lL}$ and $Z_{lR}$ scale
dependent.  However as R-parity violation is small in our analysis, the mixing
matrices are scale independent to a very good approximation.  For numerical
convenience, we rotate to a diagonal charged lepton Yukawa matrix basis at
$\MX$ before renormalizing 
down to $\MZ$.  In general this means that there are a lot of LNV parameters
in the diagonal basis, however they are all related to each other by the rotation. 

We will first investigate the case with one non-zero dimensionless LNV coupling at
$\MX$, and then go on to discuss the case of 2 non-zero LNV parameters.  
The RG evolution will dynamically generate all LNV parameters allowed by
symmetry. This includes the dimensionful $\mu_i$'s, which implies non-zero
sneutrino vacuum expectation values (vev) upon minimizing the Higgs-sneutrino potential \cite{0309196}.  
These effects
are important in determining the neutrino mass pattern and will be
described in detail in the following sections.  The mass matrices used to
compute the radiative corrections of the neutrinos and neutralinos are
presented in Appendix~\ref{Appendix: mass matrices} and references therein. 


The tree level, $7\times 7$ neutrino-neutralino mass matrix $\mathcal{M}_N$
generates an effective $3\times3$ neutrino mass matrix through a seesaw
mechanism.  Writing $\mM_N$ as defined in Appendix~\ref{Appendix: mass
  matrices}, 
\eqa
\mM_N &=& \left( \begin{array}{cc}
\mM_{\chi^0} & m^T \\
m & m_{\nu} \end{array} \right),
\qea
where $m_{\nu}$ and $m$ contain the lepton number violating contributions, and $\mM_{\chi^0}$ is the neutralino mass matrix which plays the suppression role of the right handed heavy neutrinos in the standard see-saw mechanism \cite{Ramond}.
  The effective mass matrix $\mM^{\nu}_{\textrm{eff}}$ is then given by
\eqa \label{eq:seesaw}
\mM^{\nu}_{\textrm{eff}} &=& m_{\nu} - m \mM_{\chi^0}^{-1} m^T.
\qea
At tree level, $m_{\nu}=0_{3\times 3}$ and the effective mass matrix is given by \cite{9508271, 9408224}
\eqa
\mM^{\nu}_{\textrm{eff}} &=& \frac{(M_1g^2_2 + M_2g^2)}{2 \mu [v_u v_d(M_1g^2_2 + M_2g^2)-\mu M_1M_2]} 
\left( \begin{array}{ccc}
\Lambda_e\Lambda_e & \Lambda_e\Lambda_{\mu} &\Lambda_e\Lambda_{\tau}\\
\Lambda_{\mu}\Lambda_e & \Lambda_{\mu}\Lambda_{\mu} &\Lambda_{\mu}\Lambda_{\tau}\\
\Lambda_{\tau}\Lambda_e & \Lambda_{\tau}\Lambda_{\mu} &\Lambda_{\tau}\Lambda_{\tau} \end{array} \right).
\qea
$g_2,g$ are the SU(2) and U(1)$_Y$ gauge couplings, and $v_u,
v_d$ are the vacuum expectation values (vevs) of the neutral components of $H_u, H_d$ respectively. Also,
\eqa \label{eq:tree lambda}
\Lambda_i &\equiv& \mu v_i - v_d \mu_i, \qquad i=\{e,\mu,\tau\},
\qea
and so $\mM^{\nu}_{\textrm{eff}}$
 depends on the sneutrino vevs $v_i$ and LNV supersymmetric bilinear
 parameters $\mu_i$.  $\mM^{\nu}_{\textrm{eff}}$ is of rank 1, so there
 is one non-zero eigenvalue 
\eqa
m_{\textrm{heavy}} &=& \frac{(M_1g^2_2 + M_2g^2)\sum_i \Lambda^2_i}{2\mu[v_u v_d(M_1g^2_2 + M_2g^2)-\mu M_1M_2]}.
\qea

Clearly this is not compatible with the experimentally measured mass squares differences, which require at least two non-zero neutrino masses.  Including one-loop contributions can give mass to the other neutrinos.  In the rest of this work we will present a minimal construction in which the observed mass pattern can be obtained.

The complete one-loop self-energies for the neutralinos and neutrinos in 2
component notation, including all terms with R-parity violation are given in
\cite{0603225}.  Here we present the general expression in a matrix form
compatible with the notation of \cite{9606211}.  The one-loop corrected neutrino-neutralino mass matrix $\mathcal{M}'_N$ is given by
\eqa
(\mathcal{M}'_N)_{ij} &=& (\mathcal{M}_N)_{ij} + \frac{1}{2}(\delta\mM + \delta\mM^T)_{ij},
\qea
where
\eqa \label{wfwf}
\delta \mM_{ij} &=& (\Sigma_D)_{ij} - (\mM_N)_{ik}(\Sigma_L)_{kj},
\qea
and $\Sigma_D$ and $\Sigma_L$ are mass corrections and wavefunction renormalization respectively.

As in \cite{9606211}, we perform the full renormalization with external legs
in a flavour basis.  The Feynman rules including all LNV terms can be obtained
in the Appendix of \cite{0603225}, and by rotating one external neutral
fermion leg from mass basis into flavour basis.  The physical masses are
obtained by diagonalising $\mM'_N$.

Upon diagonalising $\mM'_N$, a $7\times 7$ mixing matrix $Z'_N$ is obtained.
As at tree level, one may also obtain an effective neutrino mass matrix
$\mM'^{\nu}_{\textrm{eff}}$ using Eq.~(\ref{eq:seesaw}).  In particular, the $3\times 3$ sub-matrix $m_{\nu}$ receives radiative contributions and plays a significant role in determining the effective neutrino mass matrix.  The mixing between the neutrinos and the
neutralinos is suppressed by the smallness of the LNV couplings and $v_i$ with respect to the
other neutralino matrix parameters, and is neglected for the purpose of calculating
the PMNS matrix.  The neutrino mixing matrix $Z_{\nu}$ is then defined by
\eq
Z_{\nu}^T \mM'^{\nu}_{\textrm{eff}} Z_{\nu} = \textrm{diag}[m_i], \qquad
i=1,2,3. 
\qe

The charged lepton-chargino mass matrix is treated similarly, where one can
again use a see-saw mechanism to obtain an effective mass matrix with the
charginos providing the see-saw suppression.  In this case however, the see
saw contribution to the effective mass matrix is suppressed by two powers of
LNV parameters, and is negligible compared with $Y_E$.  So for the purpose of
calculating the charged lepton mixing matrix, one can simply use $Y_E$ and
hence the same $Z_{lL}$ and $Z_{lR}$ as before.  After rotating to a basis
with diagonal lepton Yukawa couplings at $\MX$, the lepton \emph{flavour}
changing terms in the Yukawa matrix $Y_E$ are generated by terms at least quadratic
in $\lambda$ by the RG evolution, and again such contributions are neglected
when calculating the PMNS matrix. 

The observable PMNS mixing matrix which connects the charged lepton and
neutrino mass eigenstates is defined by\footnote{Note in 4 component Dirac
  notation, the $Y_E$ is typically associated with the term $\bar{L}_L Y_E E_R
  h_d$ in other conventions.  This $Y_E$ is the \emph{complex conjugate} of
  the $Y_E$ that appears in the superpotential in Eq.~(\ref{eq:superpotentials}).
  This is why we have $U_{PMNS}= Z_{lL}^T Z_{\nu}$
  instead of $U_{PMNS}=Z_{lL}^{\dagger} Z_{\nu}$.  However our numerical
  calculation assumes CP conservation, so the two are equivalent.  The results
  presented in this paper follow the convention defined by Eq.~(\ref{eq:superpotentials}), and we keep track of the complex conjugations for book
  keeping purposes.} 
\eqa \label{eq:PMNS}
U_{PMNS} &=& Z_{lL}^T Z_{\nu},
\qea
where $Z_{lL}$ is the mixing matrix of the left handed charged leptons.  
In the present paper, we always work in the CP-conserving limit and thus
$Z_{1L}^\dagger=Z_{1L}^T$. 
Because we are already working in a basis where the charged lepton mass matrix
is practically diagonal, the PMNS matrix is the same as $Z_{\nu}$, and they
will be used inter-changeably in the rest of the paper. 

\subsection*{The ratio of tree level to loop level mass scales}
It is well known that even if the high scale theory has all bilinear LNV
couplings set to zero, the RG flow will dynamically generate $\mu_i$,
$\tilde{D}_i$ and $m^2_{H_d L_i}$ at low scale, and hence $v_i$ after
minimizing the higgs-sneutrino scalar potential.  These effects are important,
because the dimensionless LNV couplings contribute to the effective neutrino
mass matrix at both loop level and tree level in a related way.  Typically the tree
level mass scale dominates over the loop contributions, despite the fact that
the bilinear parameters themselves are originated from radiative corrections.
Also, the loop induced mass matrix tends to align with (i.e.\ is proportional
to)
the tree level mass
matrix, further suppressing radiatively induced neutrino masses. 

\FIGURE{
\scalebox{1.0}{
  \begin{picture}(210,75)(0,0)
    \ArrowLine(10,35)(60,35)
    \ArrowArc(80,35)(20,0,180)
    \ArrowArcn(80,35)(20,0,180)
    \ArrowLine(100,35)(150,35)
    \ArrowLine(200,35)(150,35)
    \Vertex(60,35){1.5}
    \Vertex(100,35){1.5}
    \Vertex(150,35){1.5} 
    \put(35,40){$L_i$}
    \put(45,25){$\lambda'_{iaq}$}
    \put(80,60){$Q_a$}
    \put(80,5){$D^c_q$}
    \put(101,24){$(Y_D^*)_{aq}$}
    \put(120,40){$H_D$}
    \put(150,25){$\mu$}
    \put(175,40){$H_U$}
  \end{picture}
}
  \label{fig:dynamical generation of kappa}
  \caption{Dynamical generation of $\mu_i$.}
}

With only one trilinear LNV parameter in the diagonal charged
lepton basis, the above statements are generally true.  However one LNV
parameter in the weak interaction basis corresponds to many LNV couplings in a
diagonal charged lepton basis, related by the charged lepton mixings.  These
rotations need not be small because the observed PMNS mixings are near tri-bi
maximal, so we may expect a `democratic' distribution of LNV couplings.
However the fact that these parameters are related implies that 
both the alignment effect and the hierarchy between the tree and loop mass
scale persists.  We will show how this happens in the following, and then go
on to discuss the next minimal case of two LNV operators (in
the weak interaction basis). 

In scenarios with R-parity violation dominated by one $\lambda$, one can
approximate the tree level mass scale by \cite{0309196} (see also
\cite{9610540})\footnote{Note that the normalization of the higgs vev in the current paper is different from \cite{0309196}.  See Appendix~\ref{Appendix: mass matrices}.}
\eq \label{eq:tree scale approx}
m^{tree}_{\nu} \simeq -\frac{8\pi\alpha_{GUT}}{5M_{1/2}} [\frac{v_d}{16\pi^2}]^2 [\textrm{ln}\frac{\MX}{M_Z}]^2 [\lambda_{ijq}(Y_E^*)_{jq}]^2 f^2(\frac{\mu^2}{\MO^2};\frac{A^2_0}{\MO^2};\frac{\tilde{B}}{\MO^2};\tanb),
\qe
where $f$ is a dimensionless function of $\mathcal{O}(10)$.  To understand the
form of this expression, note that the tree level neutrino mass is obtained by
$\Lambda_i$ in Eq.~(\ref{eq:tree lambda}), which is proportional to $\mu_i$
and the sneutrino vevs $v_i$, while the latter also
scale like $\mu_i$ up to effects from the Yukawa couplings.  The dynamical
generation of $\mu_i$, neglecting sub-dominant terms proportional to
$\mu_i$ itself, is determined by the RG flow 
\eqa \label{eq:dyn gen}
16\pi^2 \frac{d}{dt}\mu_i &=& -\mu(\lambda_{ijk}(Y_E^*)_{jk}+3\lambda'_{ijk}(Y_D^*)_{jk}),
\qea
so a first estimate of $\mu_i$ at $\MZ$ may be obtained by naively integrating the right hand side of the above equation, yielding
\eqa
\mu_i &=&
\frac{1}{(4\pi)^2}\textrm{ln}(\frac{\MX}{\MZ})(\lambda_{ijk}(Y_E^*)_{jk}+3\lambda'_{ijk}(Y_D^*)_{jk})\mu
+ \mathcal{O} \left(\frac{1}{(4 \pi)^4}\ln^2\frac{M_X}{M_Z}\right).\label{heehee}
\qea
A pseudo-Feynman diagram in terms of superfields for the second term of
Eq.~(\ref{heehee}) is shown in Fig.~\ref{fig:dynamical generation of kappa}.

On the other hand, the scale of the radiative corrections can be approximated
by the part of $\Sigma_D$ in Eq.~(\ref{wfwf}) which correspond to $m_{\nu}$ in Eq.~(\ref{eq:seesaw}) (see also \cite{0603225}).
  We define this scale to be
\eq 
\label{eq:loop scale approx}
m^{loop}_{\nu}\equiv \sum_{i=1}^3\Big({\Sigma_D}_{ii}\Big)
 \simeq \sum_{i,j,k}\lambda_{ijk}\lambda_{ikj}\frac{m^*_{f_j}}{(4\pi)^2}\frac{(\mM^{2*}_{\tilde{f}_{LR}})_{kk}}{(\mM^2_{\tilde{f}_{LL}})_{kk}-(\mM^2_{\tilde{f}_{RR}})_{kk}} \textrm{ln}\frac{(\mM^2_{\tilde{f}_{LL}})_{kk}}{(\mM^2_{\tilde{f}_{RR}})_{kk}},
\qe
where the summation $i$ is over external neutrino flavour eigenstates in the basis 
defined in Appendix~\ref{Appendix: mass matrices}.  
$m^*_{f_j}$\footnote{The phases of the charged lepton mass matrix $m_f$
  can always be rotated away by redefinition of the lepton fields.  However we
  find leaving the redundant complex conjugations helpful when considering
  chirality flips in loop diagrams.} is the chirality flip of the fermion
propagator with legs labelled by $j$, and the $\mM^2$'s represent
mass insertions with appropriate scalar `handedness'.  A mass insertion diagram
 of $m_{\nu}^{loop}$ is displayed in Fig.~\ref{fig:SigmaD MIA}.
The approximation for $\lambda'$ is exactly the same, apart from extra colour factors $n_{c} =3$
for both Eq.~(\ref{eq:tree scale approx}) and Eq.~(\ref{eq:loop scale
  approx}).

The ratio $m^{tree}_{\nu}/m^{loop}_{\nu}$ can then be approximated to be
\eqa
\label{eq:tree mass approx}
\hspace*{-1.0cm}\frac{m^{tree}_{\nu}}{m^{loop}_{\nu}} &\simeq&
-n_{c}\frac{\alpha_{GUT}\textrm{ln}^2(\MX/M_{Z})}{5\pi M_{1/2}(A_0 -
  \mu\tanb)}\frac{(\mM^2_{\tilde{f}_{LL}})-(\mM^2_{\tilde{f}_{RR}})}{\textrm{ln}((\mM^2_{\tilde{f}_{LL}})/(\mM^2_{\tilde{f}_{RR}}))}
 f^2(\frac{\mu^2}{\MO^2};\frac{A^2_0}{\MO^2};\frac{\tilde{B}}{\MO^2};\tanb), \label{eq:TreeLoopEstimateLam}
\qea

where
we have made the approximation
\eq 
(\mM^2_{\tilde{f}_{LR}})_{kk}\sim m_{f_k}(A_0-\mu\tanb),
\qe
where $\tilde{f}$ represents sleptons or squarks, as appropriate.  
Setting
$\alpha_{GUT} = 0.041$ and $\MX = 10^{16}$ GeV, a rough scan along the SPS1
slope \cite{SPS1a} with $\Mhalf=[250,750]$ and also $\tanb=[10,40]$ gives
$m^{tree}_{\nu}/m^{loop}_{\nu}$ of order $\mathcal{O}(30-80)$ for
$\lam_{ijj}$, and $\mathcal{O}(70-200)$ for $\lam'_{ijj}$, in agreement with
the approximation of Eq.~(\ref{eq:tree mass approx}).  However, this is too
large to account for the mild 
empirical neutrino mass hierarchy of a factor of $\lsim$6 from
Eq.~(\ref{eq:neutrino_data}), assuming negligible lightest neutrino mass\footnote{In the
  opposite extreme of a quasi-degenerate mass spectrum, there is clearly no
  mass `hierarchy'.}. 

Note that in the above discussion, alignment (proportionality) between the
tree level neutrino effective mass matrix and 
the radiative correction has not yet been included. The inclusion of such
effects will enhance the
hierarchy between the heavy and the light neutrino masses obtained in
the above approximation.  
Because at tree level there is only one massive
neutrino, exact alignment would imply still only one massive neutrino 
after the inclusion of radiative corrections.  
Misalignment between the tree level
contributions and the radiative corrections will give rise to light neutrino
masses.  The size of the mass prediction for the second lightest neutrino
compared to the heaviest neutrino, and hence $|\Delta m_{21}^2|$ compared
 to $|\Delta m_{31}^2|$, can be estimated by the
amount of misalignment, multiplied by the ratio of the loop/tree level
scales.
\FIGURE{
  \begin{picture}(210,75)(0,0)
    \DashArrowArc(105,40)(20,90,180){3}
    \DashArrowArcn(105,40)(20,90,0){3}
    \ArrowArcn(105,40)(20,270,180)
    \ArrowArc(105,40)(20,270,360)
    \ArrowLine(45,40)(85,40)
    \ArrowLine(165,40)(125,40)
    \Vertex(105,20){1.5}
    \Vertex(105,60){1.5}
    \put(35,39){$\nu_i$}
    \put(167,39){$\nu_i$}
    \put(60,28){$\lambda_{ijk}$}
    \put(128,28){$\lambda_{ikj}$}
    \put(98,65){$(\mM^{2*}_{\tilde{f}_{LR}})_{kk}$}
    \put(100,10){$m^*_{f_j}$}
  \end{picture}
  \caption{A mass insertion diagram which represents the loop corrections to $m_{\nu}^{loop}$ by $\lambda_{ijk}$ and $\lambda_{ikj}$.}
  \label{fig:SigmaD MIA}
}

To discuss possible alignment effects, it is instructive
 to first start in the charged lepton mass basis with one
LNV coupling.  In the latter case, if the LNV coupling violates only one lepton flavour,
 then the other two lepton flavours are separately conserved.  Each of the tree
 level and loop level neutrino mass matrices can only have one non-zero entry corresponding
 to the lepton flavour violated, and so these contributions to the full
 neutrino mass matrix are exactly aligned.  Using this flavour symmetry
 argument it is easy to see that this alignment will persist to all
 orders in perturbation theory.  On the other hand, if the LNV coupling violates
all three lepton flavours, then no bilinear LNV couplings can be dynamically
generated.  
There is no source which 
violates lepton flavour by two units, therefore we cannot generate a
 one-flavour 
 violating bilinear term from a three-flavour violating term.
This can be seen explicitly in the
one-loop approximation by reference to Eq.~(\ref{eq:dyn gen}).
Thus no tree level mass matrix can be 
generated.  The same reasoning, together with the requirement of an odd number
of chirality flips in the loop diagrams, also implies that the
radiative correction to the neutrino mass matrix is zero. 

When the one LNV coupling is defined in the weak interaction basis, which need
not coincide with the diagonal charged lepton basis, exact alignment will
cease to hold, as the above symmetry arguments do not apply any more.  However,
it is possible to show that to lowest order in perturbation theory, there is
still alignment between the tree level and loop level contributions.  For this
purpose, we work in a diagonal lepton basis, and make use of the fact that in
the R-parity conserving limit, the mixing matrices that rotate into the
diagonal basis obey
\eq
Z_{lL}=Z_{\tilde{l}L} \qquad Z_{lR}=Z_{\tilde{l}R} \qquad
(\mathcal{M}^2_{\tilde{l}_{LR}})_{jk}\propto (Y_E)_{jk}. \label{diagB}
\qe
Eq.~(\ref{diagB}) is true at all renormalization scales, as discussed in the
previous section with universal boundary conditions due to the absence of
intrinsic lepton flavour violation.  
The universality of the lepton and slepton
mixings will be violated at order $\lambda^2$, 
but we shall study small values ($\lsim 10^{-3}$) of $\lambda$ and so
the non-universality is neglected in our lowest
order approximation.  We will also make use of the fact that all
 dynamically generated LNV couplings are small, as will be seen shortly.  

In the following we denote parameters in the diagonal lepton basis with a hat.
Although we use $\hat{\lambda}$ for illustration, the case of $\hat{\lambda}'$
is similar.  The dimensionless LNV couplings in the diagonal lepton basis,
$\hat{\lambda}_{abc}$ are related to the LNV couplings in the weak interaction
basis, $\lambda_{ijk}$, by 
\eq
\hat{\lambda}_{abc}=\lambda_{ijk}(Z^{\dagger}_{lL})_{ai}(Z^{\dagger}_{lL})_{bj}(Z_{lR})_{kc}.
\qe
The tree level neutrino mass matrix then scales (neglecting the sneutrino
vevs) as 
\eqa\label{eq:tree MIA}
(\hat{\mM}^{\nu}_{\textrm{eff}})_{ab} &\propto& \hat{\Lambda}_a\hat{\Lambda}_b \propto \hat{\mu}_a\hat{\mu}_b \propto \hat{\lambda}_{acd}(\hat{Y}_E^*)_{cd}\hat{\lambda}_{bef}(\hat{Y}_E^*)_{ef} \nonumber \\
&=& \Big[\lambda_{ijk}(Y_E^*)_{jk}\lambda_{nlm}(Y_E^*)_{lm}\Big](Z^{\dagger}_{lL})_{ai}(Z^*_{lL})_{nb},
\qea
and so
\vspace{-0.1cm}
\eq
(\hat{\mM}^{\nu}_{\textrm{eff}})_{ab}=(Z^{\dagger}_{lL})_{ai}(\mM^{\nu}_{\textrm{eff}})_{in}(Z^*_{lL})_{nb},
\qe
where $\mM^{\nu}_{\textrm{eff}}$ is the effective neutrino mass matrix in the original lepton flavour basis.  The factorization of the two charged lepton mixings represents the shift of $Z_{lL}$ from the charged lepton sector to the neutrino sector when computing $U_{PMNS}$ using Eq.~(\ref{eq:PMNS}).  At loop level, we approximate its contribution to the full mass matrix by the dominating effect from $\hat{\Sigma}_D$
\eq
(\hat{\Sigma}_D^{\nu})_{ab} \simeq -\sum_{c,f}\sum_{d,e}\biggl\{\frac{(\hat{m}^*_l)_{cf}}{(4\pi)^2}\hat{\lambda}_{acd}\hat{\lambda}_{bef}B_0(0,\hat{m}^2_{\tilde{l}_{de}},\hat{m}^2_{l_{cf}})+ (a\leftrightarrow b)\biggr\}
\qe
in self-evident notation, and where $B_0$ is a Passarino-Veltman function
\cite{PV}.  Using a mass insertion approximation (MIA), one obtains
\eqa\label{eq:sigmaD MIA}
(\hat{\Sigma}^{\nu}_D)_{ab} &\simeq & \sum_{c,f}\sum_{d,e}\biggl\{\frac{(\hat{m}^*_l)_{cf}}{(4\pi)^2}\frac{\hat{\lambda}_{acd}\hat{\lambda}_{bef}(\hat{\mM}^{2*}_{\tilde{l}_{LR}})_{ed}}{(\hat{\mM}^2_{\tilde{l}_{LL}})_{ee}-(\hat{\mM}^2_{\tilde{l}_{RR}})_{dd}}\textrm{ln}\Big(\frac{(\hat{\mM}^2_{\tilde{l}_{LL}})_{ee}}{(\hat{\mM}^2_{\tilde{l}_{RR}})_{dd}}\Big) + (a\leftrightarrow b)\biggr\}\nonumber \\
&=& \sum_{\substack{i,j,k\\n,l,m}}\sum_{d,e}\biggl\{\frac{(m^*_l)_{jm}}{(4\pi)^2}\lambda_{ijk}\lambda_{nlm} \Big[\frac{(Z^*_{lL})_{le}(\hat{\mM}^{2*}_{\tilde{l}_{LR}})_{ed}(Z^{T}_{lR})_{dk}}{(\hat{\mM}^2_{\tilde{l}_{LL}})_{ee}-(\hat{\mM}^2_{\tilde{l}_{RR}})_{dd}}\textrm{ln}\Big(\frac{(\hat{\mM}^2_{\tilde{l}_{LL}})_{ee}}{(\hat{\mM}^2_{\tilde{l}_{RR}})_{dd}}\Big)\Big] \nonumber \\
&& \qquad\qquad + (i \leftrightarrow n)\biggr\}(Z^{\dagger}_{lL})_{ai}(Z^*_{lL})_{nb} \nonumber \\
&=& \sum_{\substack{i,j,k\\n,l,m}}\biggl\{\frac{1}{(4\pi)^2}\lambda_{ijk}\lambda_{nlm} \frac{(m^*_l)_{jm}(\mM^{2*}_{\tilde{l}_{LR}})_{lk}}{\overline{\mM}^2_{\tilde{l}_{LL}}-\overline{\mM}^2_{\tilde{l}_{RR}}} \textrm{ln}\Big(\frac{(\overline{\mM}^2_{\tilde{l}_{LL}})}{(\overline{\mM}^2_{\tilde{l}_{RR}})}\Big)\Big[1+\mathcal{O}\Big(\frac{\delta\mM^2}{\overline{\mM}^2}\Big)\Big] \nonumber \\
&& \qquad\quad + (i\leftrightarrow n)\biggr\}(Z^{\dagger}_{lL})_{ai}(Z^*_{lL})_{nb},
\qea
where $\overline{\mM}^2_{\tilde{l}_{LL(RR)}}$ is a mean squared mass of the left (right) handed sleptons, and $\delta\mM^2$ represents deviation from such mean values.  Note the index contraction of the $\lambda$ in Eq.~(\ref{eq:sigmaD MIA}) is different from those in Eq.~(\ref{eq:tree MIA}).  However $\hat{\Sigma}_D^{\nu}$ is dominated by the one LNV coupling that is non-zero at $\MX$, as all dynamically generated couplings are small compared with it.  We then have $i=n$, $j=l$ and $k=m$ and consequently $\hat{\Sigma}_{D}^{\nu}$ is proportional to the tree level effective mass matrix $\hat{\mM}^{\nu}_{\textrm{eff}}$, i.e.
\eq
(\hat{\Sigma}_{D}^{\nu})_{ab} \propto (\hat{\mM}^{\nu}_{\textrm{eff}})_{ab},
\qe
up to higher order corrections.  It is unwieldy to write down the higher order
corrections to this expression, with terms from many sources (e.g. RGE, mixing
matrices including the charginos-lepton mixings and corrections in MIA)
entering 
the next lowest order perturbation.  The key result here is that the index
contractions in Eq.~(\ref{eq:tree MIA}) and Eq.~(\ref{eq:sigmaD MIA}) are
different, unless they are dominated by a single LNV parameter, in which case
they become proportional to each other.  The tree level mass
 matrix is bilinear in $\Lambda_i$, which is proportional to $\mu_i$.  The
 dynamical generation of $\mu_i$ can be obtained by integrating their
 RGEs, which contain terms with odd powers of $\lam_{ijk}$ and $\lam'_{ijk}$.
  On the other hand, one obtains the loop corrections to the neutrino mass
 matrix by summing over contributions which contain even powers of the LNV
 couplings.  The interference between various contributions at tree level
 and at loop level are in general different as a result.

The scale of the light neutrino mass can be estimated by noting that
 in the case of degenerate charged lepton masses, there is an ambiguity
 in the definition of the charged lepton mixing.  The Yukawa matrix is simply a unit matrix, which implies that our
 earlier discussion on one LNV coupling in a diagonal charged lepton basis
 will apply in this situation, and so there is only one massive neutrino.  In
 other words, the light neutrino mass $m_{\nu}^{\textrm{light}}$ arises from the deviation of charged lepton universality.  Its mass is therefore expected
 to be suppressed relative to the heavy neutrino mass $m_{\nu}^{\textrm{heavy}}$
 by (the square of) the ratio of the charged lepton mass to the scalar
 mass scale that appears in the loop, as well as the tree-loop scale
 difference coming from the RG evolution, ie
\eqa
\frac{m_{\nu}^{\textrm{light}}}{m_{\nu}^{\textrm{heavy}}}&\sim&\mathcal{O}\Big((\frac{m_l}{m_{\tilde{l}}})^2\frac{1}{f^2}\Big),
\qea
with $f= f(\frac{\mu^2}{\MO^2};\frac{A^2_0}{\MO^2};\frac{\tilde{B}}{\MO^2};\tanb)$ the function from Eq.~(\ref{eq:tree mass approx}).

With more than one dominant LNV coupling, the situation is very different.  By
choosing appropriate LNV parameters, one can obtain a partial cancellation
between the dimensionless LNV contributions in the dynamical generation of
$\mu_i$, and suppress the tree level neutrino mass as a result.  Also the
above derivation of alignment relies on one coupling dominance, specifically
$i=n$, $j=l$ and $k=m$ Eq.~(\ref{eq:sigmaD MIA}).  Once we have more than
 one dominant coupling, this
manipulation ceases to hold.  As the alignment effect is weakened and the mass
hierarchy becomes smaller, larger PMNS mixing is expected with large
charged lepton mixing angles.  So in the next minimal case with 2 LNV
couplings at $\MX$, by allowing partial cancellations in the $\mu_i$
generation and large charged lepton mixings, we may already be able to obtain
the observed neutrino mass pattern.  We will investigate this possibility by a
numerical approach, discussed in the next section. 

\section{Numerical procedure}\label{sec:numerics}

The parameter set that defines our model is given by
\eqa \label{eq:parameter set}
\MO, \Mhalf, \AO, \sgnmu,  &@& \MX, \nonumber \\
\theta^{l}_{12},\theta^{l}_{13},\theta^{l}_{23}, \Lambda_1, \Lambda_2 &@& \MX,\nonumber \\
 \tanb &@& \MZ, \qea
where
\eq
\Lambda_1,\Lambda_2 \in \{\lambda_{ijk},\lambda'_{ijk}\},
\qe
and $\MX$ is defined to be the scale at which the electroweak gauge couplings unify.  
$\MO$, $\Mhalf$ and $\AO$ are the universal scalar mass, gaugino mass and
SUSY breaking trilinear scalar coupling (divided by the corresponding Yukawa
coupling) at $\MX$
 respectively.  The sign of $\mu$ is defined by $\sgnmu$.
 The
 charged lepton rotations are characterized by $\theta^{l}$'s in the 
standard parameterization \cite{PDG}, and $\tanb$ is the ratio of the higgs vacuum 
expectation values $v_u/v_d$.

In the quark sector, flavour mixing is assumed to reside in the down-quark
sector at the weak scale.  We have also assumed that the down-quark Yukawa matrix $Y_D$ is symmetric at that scale.
  In the lepton sector, we work with non-diagonal charged lepton Yukawa
matrices, and define any non-zero LNV parameters in that flavour basis.  The
mixing angles $\theta^{l}$ are determined by fitting the PMNS matrix.  For definiteness,
we assume  that the two charged lepton mixings $Z_{lL}$ and $Z_{lR}$ are the same,
 or equivalently, that the charged lepton Yukawa matrix $Y_E$ is symmetric.

The set of Lagrangian parameters (RPC and LNV) at $\MZ$ are obtained using a
modified version of the \softsusy~code~\cite{softsusy}.  At $\MX$, a set of
dimensionless LNV parameters defined in Eq.~(\ref{eq:parameter set}) is
chosen.  At $\MZ$, $\tanb$, the ratio of higgs vevs is also specified.  The
high scale LNV parameters are rotated to a diagonal charged lepton basis
specified by $\theta^l$.

The parameters are run to $\MZ$ using the full set of 1-loop MSSM RGEs
including all LNV contributions, which are then required to match with the low
scale boundary conditions on the gauge and Yukawa couplings.  Throughout the analysis,
 we use $m_t = 172.7$ GeV \cite{ex0507091} and $\MZ=91.1876$ GeV for the pole masses of the top quark 
and the $Z^0$ boson respectively.  The weak scale gauge couplings in the $\overline{MS}$ scheme
 are set to be $\alpha^{-1}(\MZ)=127.918$ and $\alpha_s(\MZ)=0.1187$.  Light quark masses are
 also set to their central values in the $\overline{MS}$ scheme: $m_b(m_b)=4.25$ GeV, $m_c(m_c)=1.2$ GeV,
 $m_s(2 \textrm{GeV})=0.1175$ GeV, $m_d(2 \textrm{GeV})=0.00675$ GeV and $m_u(2 \textrm{GeV})=0.003$ GeV \cite{PDG}.
  For the calculation of the physical Higgs mass and electroweak symmetry breaking (EWSB) conditions, one loop corrections are included.

To calculate the neutrino masses, we run the parameter set to
\eq
M_S=\sqrt{m_{{\tilde t}_1}(M_S) m_{{\tilde t}_2}(M_S)},
\qe
where the EWSB conditions are imposed and all sparticle and Higgs pole masses
are calculated in 
\softsusy.  The neutrino masses are obtained from the full $7\times 7$
neutrino-neutralino mass matrix including all LNV contributions at the 1-loop
level.  Throughout the calculation, the (s)lepton doublets are put on the same
footing as $H_d$.  Our analytic results were checked against Ref.\cite{0603225} for
the full LNV case, and against Ref.\cite{9606211} in the RPC limit.  Our numerical results
agree well with Ref.\cite{0603225}.  In the RPC limit, the neutralino corrections
are also in agreement with the latest published version of \softsusy. 

After the initial rotation to a diagonal $Y_E$ at $\MX$, the charged lepton
matrix remains essentially diagonal at all renormalization scales due to the
smalless of the LNV couplings.  The computation of the PMNS mixings thus
neglects all `residual' flavour mixing generated through renormalization in the charged lepton sector.  The three
charged lepton mixing angles, together with the two GUT-scale LNV parameters
are varied in order to obtain the best fit.

The minimization to find the best fit parameter set is performed using MINUIT
\cite{MINUIT} and the observables assumed in Eq.~(\ref{eq:neutrino_data}).  $\chi^2$ is defined to be
\eq
\chi^2=\sum_{i=1}^{N_{obs}}\left(\frac{f_i-O_i}{\sigma_i}\right)^2, \label{eq:chi squared ratio}
\qe
where $O_i$ are the central values of the $N_{obs}$ experimental observables, $f_i$ are the corresponding
 numerical predictions, and $\sigma_i$ are the 1-sigma uncertainties.  
We find that it is difficult to obtain convergence 
of a procedure which  minimizes $\chi^2$ when
solar and atmospheric mass squared differences are included in $O_i$ directly.
We find a multi-step minimisation approach to be effective.
In this approach, we first minimize a 
 $\chi^2$ of the mass ratio of the 2 heavy neutrinos and the sine squared of
the three PMNS mixing 
 angles by keeping one LNV parameter fixed.  For a normal hierarchy we have
\eq
\begin{array}{lll}
m_{\nu_1}  \sim 0, & \quad m_{\nu_2}  =  8.9^{+0.15}_{-0.16} \times 10^{-3} \textrm{eV}, & \quad m_{\nu_3}  =  5.10\pm 0.20 \times 10^{-2} \textrm{eV},
\end{array}
\qe
\eqa
 \Rightarrow \frac{m_3}{m_2} &=&  5.74\pm0.32, \nonumber
\qea 
whereas for an inverted hierarchy, we have
\eq
\begin{array}{lll}
m_{\nu_1}  =  5.10\pm 0.20 \times 10^{-2} \textrm{eV},& \quad m_{\nu_2}  =  5.18\pm 0.19 \times 10^{-2} \textrm{eV}, & \quad m_{\nu_3}  \sim  0, \\
\end{array}
\qe
\eqa
\Rightarrow \frac{m_2}{m_1} =  1.0151\pm0.0769. \nonumber
\qea

At this stage, we obtain a good fit to the charged lepton mixing angles and the mass ratio but
 the values of the neutrino masses in general do not fit the atmospheric and solar mass squared
 differences.  To perform a `proper' fit we then switch to a $\chi^2$ of $\Delta m^2_{21}$,
 $|\Delta m^2_{31}|$ as well as the three PMNS angles.  We first fix the three charged lepton
 mixing angles and allow the two LNV parameters to vary.  The LNV parameters will then adjust their
 values to fit the mass squared differences.  Finally a full five parameter fit is performed.  It
 is also helpful to have a good initial estimate of the size of the LNV parameter that is fixed
 in the first stage of minimization.  We find that a suitable starting point is the value of
the LNV coupling that saturates the one LNV coupling bound in \cite{0309196}, derived from the
 cosmological bound $\sum_i m_{\nu_i} < 0.7$ eV at $95\%$ condidence level (CL)~\cite{astroph0605362}.

The above fitting is performed at $M_S$.  After this, the resulting set of
parameters is renormalized down to $\MZ$, where specific products of couplings
are checked against the bounds in \cite{9906209}.  We also checked against the
experimental bounds on the branching ratios of the processes $l\to l'\gamma$
\cite{hep-ex/0111030, hep-ex/0508012, hep-ex/0502032}, $B_s\to\mu^+\mu^-$
\cite{CDF_BsMuMu_bound} and $b\to s\gamma$ \cite{0609263}.  The bounds we use are
\eq\label{eq:BR bounds}
\begin{array}{rcccll}
&&BR_{exp}(\mu \to e \gamma) &<& 1.2 \times 10^{-11}\qquad & 90\%\textrm{CL},\\
&&BR_{exp}(\tau \to e \gamma) &<& 1.1 \times 10^{-7} & 90\%\textrm{CL},\\
&&BR_{exp}(\tau \to \mu \gamma) &<& 6.8 \times 10^{-8} & 90\%\textrm{CL},\\
&&BR_{exp}(B_s \to \mu^+ \mu^-) &<& 1.0 \times 10^{-7} &95\%\textrm{CL},\\
2.76\times 10^{-4} &<& BR(b \to s\gamma) &<& 4.34 \times 10^{-4}
&2\sigma,  \\
\end{array}
\qe
where $BR(b \to s\gamma)$ combines statistical and systematic experimental errors
\cite{HFAG06} together 
 with a Standard Model theoretical error \cite{GambinoBuras} by adding them in
 quadrature.

We have included a detailed calculation of $BR(l\to l'\gamma)$
 in Appendix~\ref{llgamma}.  For a discussion of contributions of different couplings to
this process, see e.g. \cite{0610406}.  The LNV contribution to $BR(B_s\to
\mu^+\mu^-)$ was calculated in \cite{9701283}.  A generalisation to include
left-right mixings of the mediating squarks and non-degenerate sparticle
masses can be found in \cite{0609263}.  The LNV contribution to $BR(b\to s\gamma)$
in a leading log approximation is presented in \cite{0004067}.

One technical aspect that should be mentioned concerning the numerical
calculation is the cancellation between the CP even (CPE) and the CP odd (CPO)
scalar contributions to the neutrino masses at loop level.  In the R-parity
conserving limit, the CPE and CPO sneutrino contributions cancel exactly.
When R-parity is violated, the higgses mix with the sneutrinos through the
sneutrino vevs and the bilinear soft terms $\tilde{D}_i$, lifting the
degeneracy of the masses of the CPE and CPO scalars.
The mixing matrices of the CPE and CPO
scalars also differ.  This degeneracy lifting is in general very small compared
with the 
sneutrino mass scale, but may have a significant impact on the neutrino masses.

Such a large cancellation is numerically unstable.  In the phenomenologically
interesting LNV parameter space,  numerical fluctuations caused by
such an instability are actually insignificant for
weak scale sneutrinos.  However away from this `good' LNV region, the second-lightest
 neutrino mass is usually highly suppressed compared to the third lightest (or the heaviest)
 neutrino as discussed in the previous section.  Its eigenvalue can be dominated by
 numerical fluctuations, and tends to spoil the minimization procedure when
 the LNV 
 parameter space away from the physically interesting region is accessed.

To get around this problem, we perform an analytic approximation commonly known as mass
 insertion approximation (MIA).  We apply this technique to calculate the deviation from exact
 cancellation in the RPC limit, instead of performing the large numerical cancellation directly.
  In this calculation, it is useful to think of the $5\times 5$ neutral scalar matrices as a
 $2\times 2$ higgs block, a $3\times 3$ sneutrino block, and a $2\times 3$ block that
 mixes the higgses and the sneutrinos in the basis $(h_u, h_d, \tilde{\nu}_e, \tilde{\nu}_{\mu}, \tilde{\nu}_{\tau})$ as follows
\eqa\label{eq:CPECPO blocking}
\Big(\mM^2_{CPE(CPO)}\Big)_{5 \times 5} &=& \left( \begin{array}{cc}
(M^2_{H(A)})_{2\times 2} & (\sigma_{H(A)})_{2\times 3} \\
(\sigma^{T}_{H(A)})_{3\times 2} & (M^2_{\tilde{\nu}})_{3\times 3}
\end{array} \right).
\qea
We take advantage of the fact that in the `t Hooft-Feynman gauge, the CPE and CPO sneutrino blocks $M^2_{\tilde{\nu}}$ are identical and can be diagonalised by the same orthogonal matrix.  The difference between the CPE and CPO contributions then come from $M^2_{H(A)}$ and $\sigma_{H(A)}$.  The CPE and CPO $M^2$'s can be separately diagonalised, leaving the off-diagonal contributions entirely in the (s)lepton number violating sector.  In this basis it is possible to obtain successive corrections to the mixing matrices as well as the mass eigenvalues, or in our case their \emph{difference} in terms of the small LNV parameters.  We refer interested readers to \cite{9703442} for a review of MIA.  For details of our specific calculation, see Appendix~\ref{CPECPO}.

\section{Results}\label{sec:results}
\begin{figure}[ht!]
\centering
\subfigure[$m_3$ and $m_2$ vs $\lam'_{233}$]{\label{fig:nfull0_lampM.eps}\includegraphics[scale=0.5]{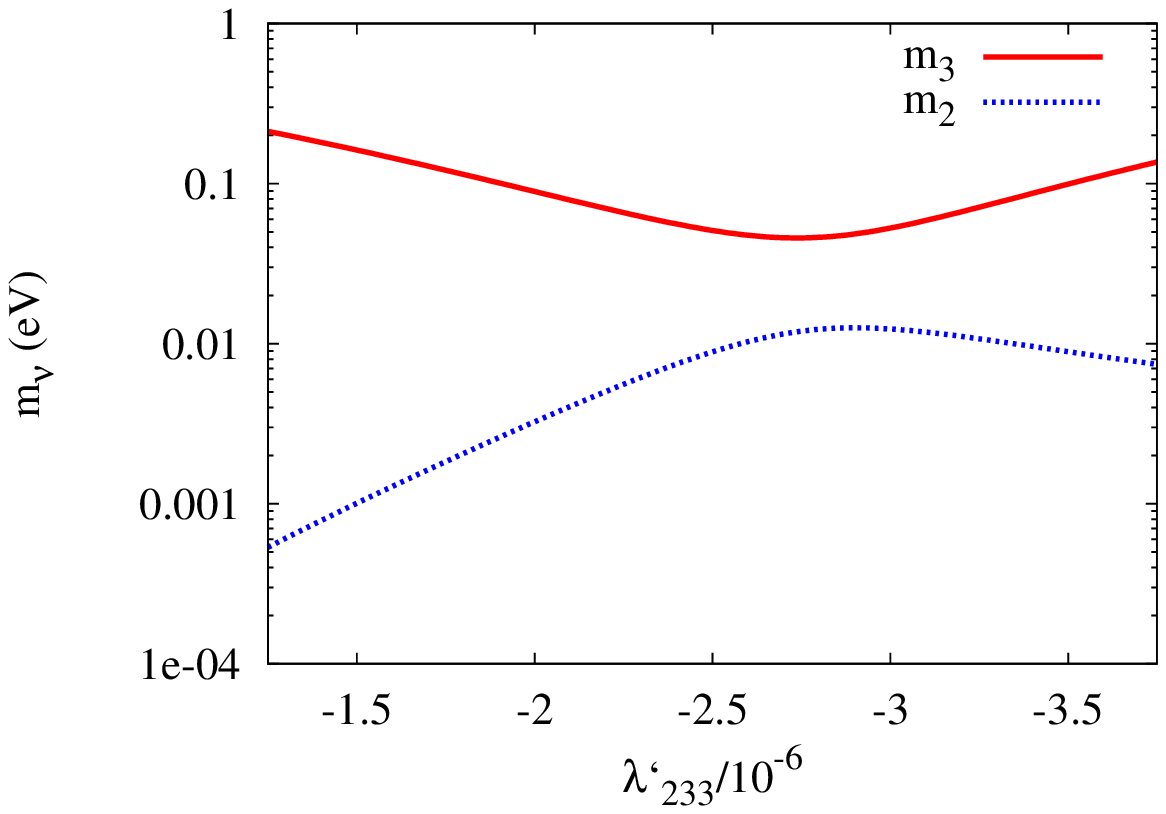} }
\subfigure[$|\Delta m^2_{31}|$ and $\Delta m^2_{21}$ vs $\lam'_{233}$]{\label{fig:nfull0_lampMM.eps}\includegraphics[scale=0.5]{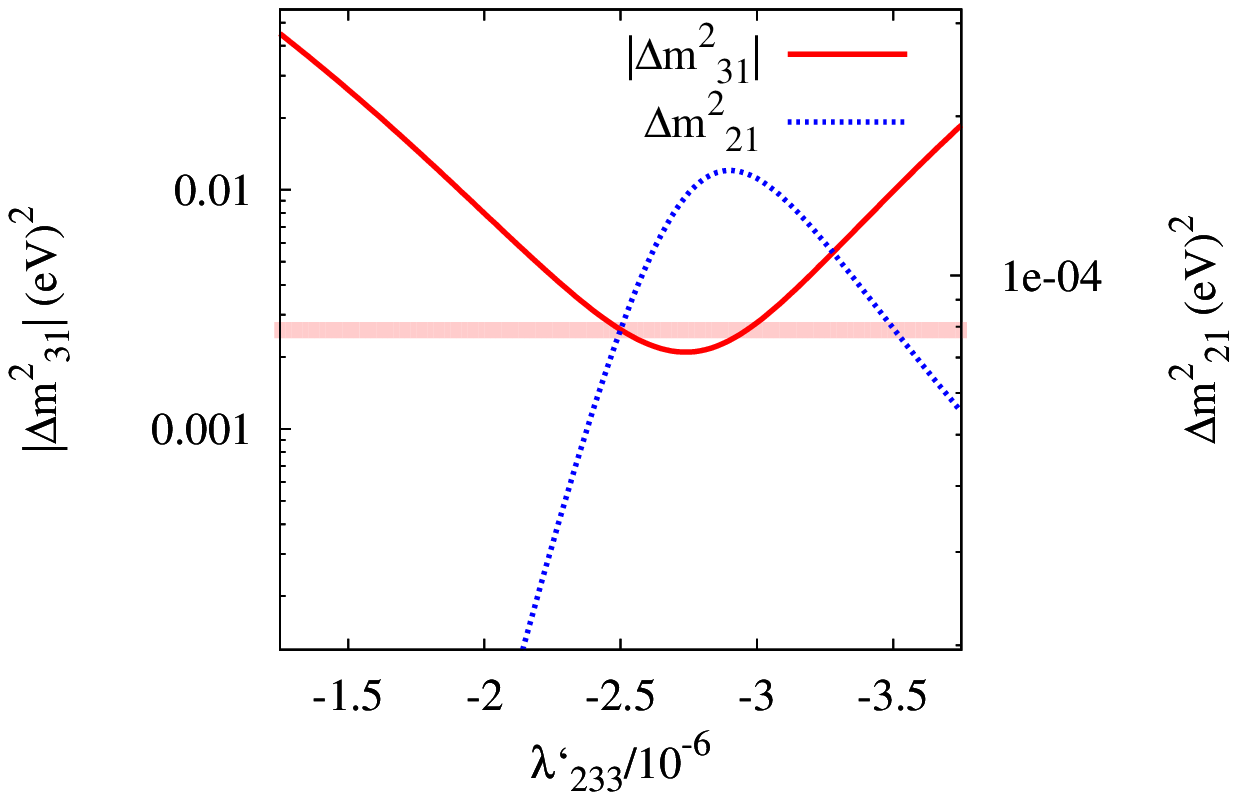} }
\subfigure[$\textrm{sin}^2\theta^{}_{13}$ vs $\lam'_{233}$]{\label{fig:nfull0_lampTheta13.eps}\includegraphics[scale=0.5]{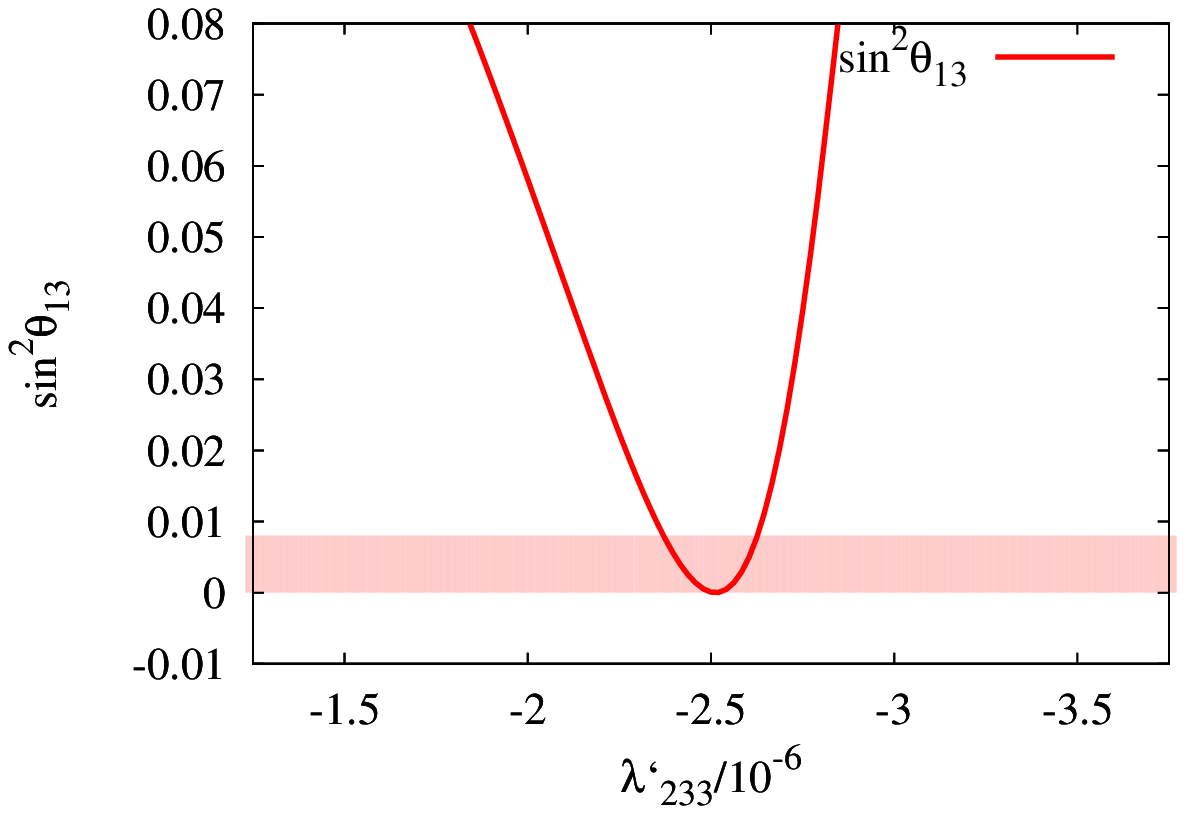} }
\subfigure[$\textrm{sin}^2\theta^{}_{12}$ and $\textrm{sin}^2\theta^{}_{23}$ vs $\lam'_{233}$]{\label{fig:nfull0_lampTheta1223.eps}\includegraphics[scale=0.5]{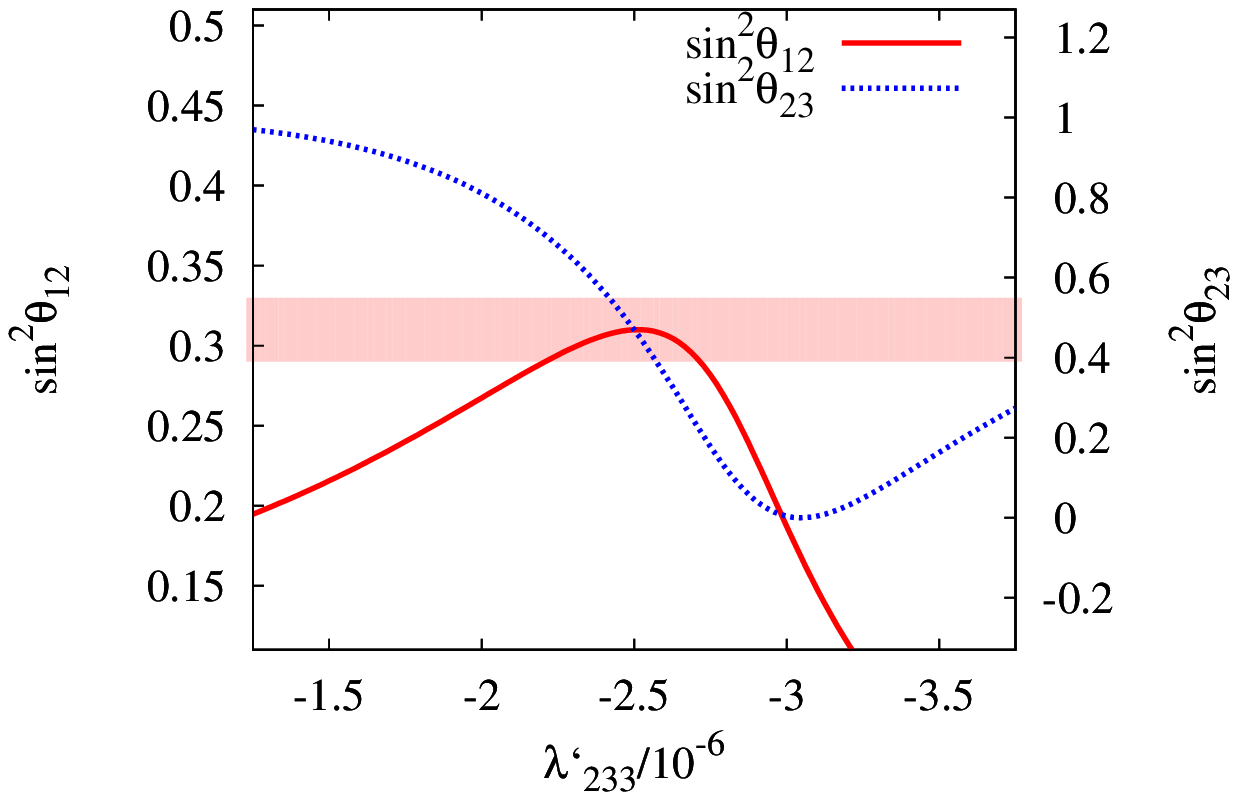} }
\caption{Variation of the observables characterising the neutrino mass pattern
  with $\lambda_{233}$ at SPS1a in a normal hierarchy.  The best fit
  parameters are $\theta^{l}_{12}=0.460$, $\theta^{l}_{13}=0.389$,
  $\theta^{l}_{23}=0.305$, $\lam_{233}=4.07\times 10^{-5}$ and
  $\lam'_{233}=-2.50\times 10^{-6}$.  The strips display the empirical $1 \sigma$
  bands.}\label{fig:normalhierarchy}
\end{figure}

For concreteness, we confine ourselves to the SPS1a \cite{SPS1a} point with
non-zero LNV couplings, i.e. 
\eq
\begin{array}{lll}
\Mhalf=250\textrm{GeV},& \quad\MO=100\textrm{GeV},& \quad\AO=-100\textrm{GeV},\\
\sgnmu = +1,& \quad\tanb =10,&
\end{array}
\qe
although the numerical procedure can be applied equally well in other mSUGRA
region.  We first display an example with a normal hierarchy with non-zero
$\lam_{233}$ and $\lam'_{233}$ at the unification scale.  The set of best fit
parameters is given by $\theta^{l}_{12}=0.460$, $\theta^{l}_{13}=0.389$,
$\theta^{l}_{23}=0.305$, $\lam_{233}=4.07\times 10^{-5}$ and
$\lam'_{233}=-2.50\times 10^{-6}$.  The variation of the observables
charaterising the neutrino mass pattern with $\lam'_{233}$ is displayed in
Fig.~\ref{fig:normalhierarchy}.  We see in Fig.~\ref{fig:nfull0_lampM.eps}
that as $\lam'_{233}$ increases, the mass difference between the two heavy
neutrinos decreases due to suppression of the tree level mass matrix and a
weakening of the alignment effect.  Eventually the mass difference increases
again, signalling a cross over where the heaviest neutrino becomes dominated
by contributions derived from $\lam'_{233}$.  In between there is a window
which allows the mass squared differences to fall within the experimentally
observed range, as displayed in Fig.~\ref{fig:nfull0_lampMM.eps}.  The
variations of $\textrm{sin}^2\theta_{12}$ and $\textrm{sin}^2\theta_{13}$
shown in  Fig.~\ref{fig:nfull0_lampTheta13.eps} and
 Fig.~\ref{fig:nfull0_lampTheta1223.eps} are associated with the cross over described
above, when the two mass eigenvectors determined by the two LNV parameters
switch. 

\begin{figure}[ht!]
\centering
\subfigure[$m_2$ and $m_1$ vs $\lam'_{233}$]{\label{fig:nifull2_lampM.eps}\includegraphics[scale=0.5]{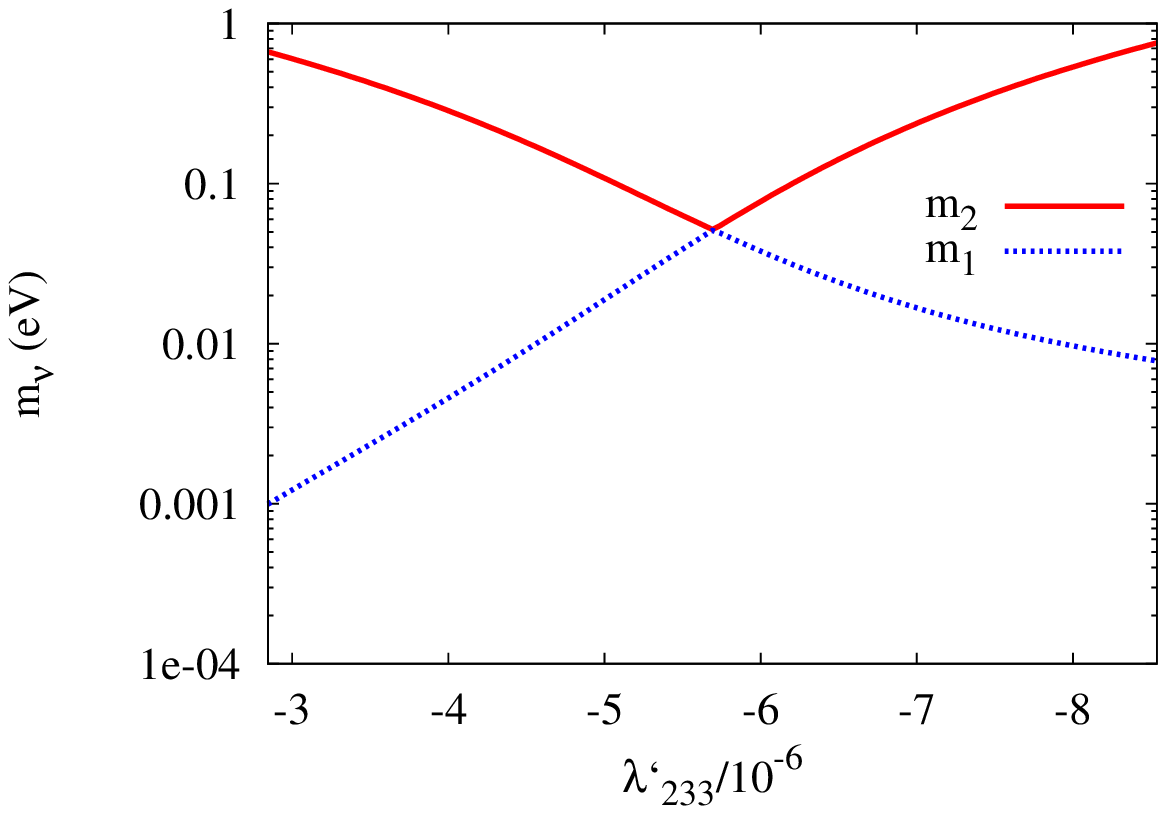} }
\subfigure[$|\Delta m^2_{31}|$ and $\Delta m^2_{21}$ vs $\lam'_{233}$]{\label{fig:nifull2_lampMM.eps}\includegraphics[scale=0.5]{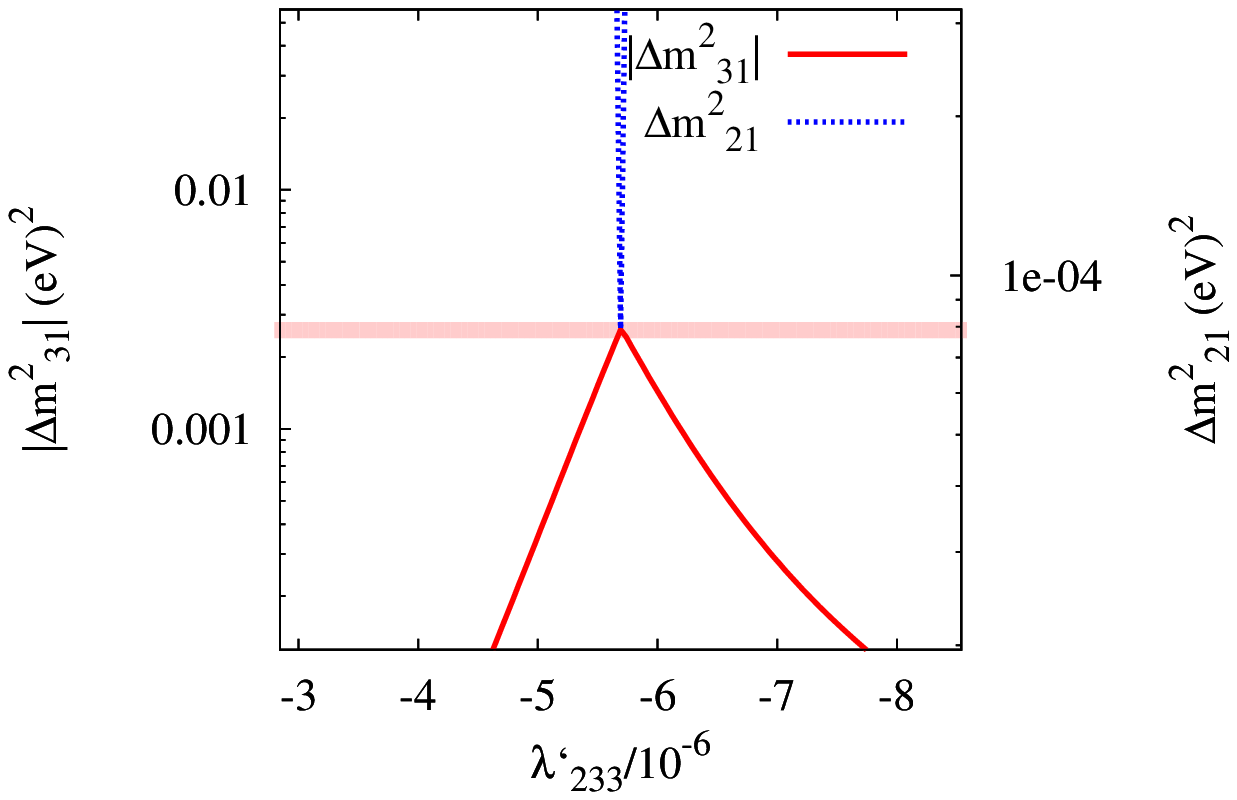} }
\subfigure[$\textrm{sin}^2\theta^{}_{13}$ vs $\lam'_{233}$]{\label{fig:nifull2_lampTheta13.eps}\includegraphics[scale=0.5]{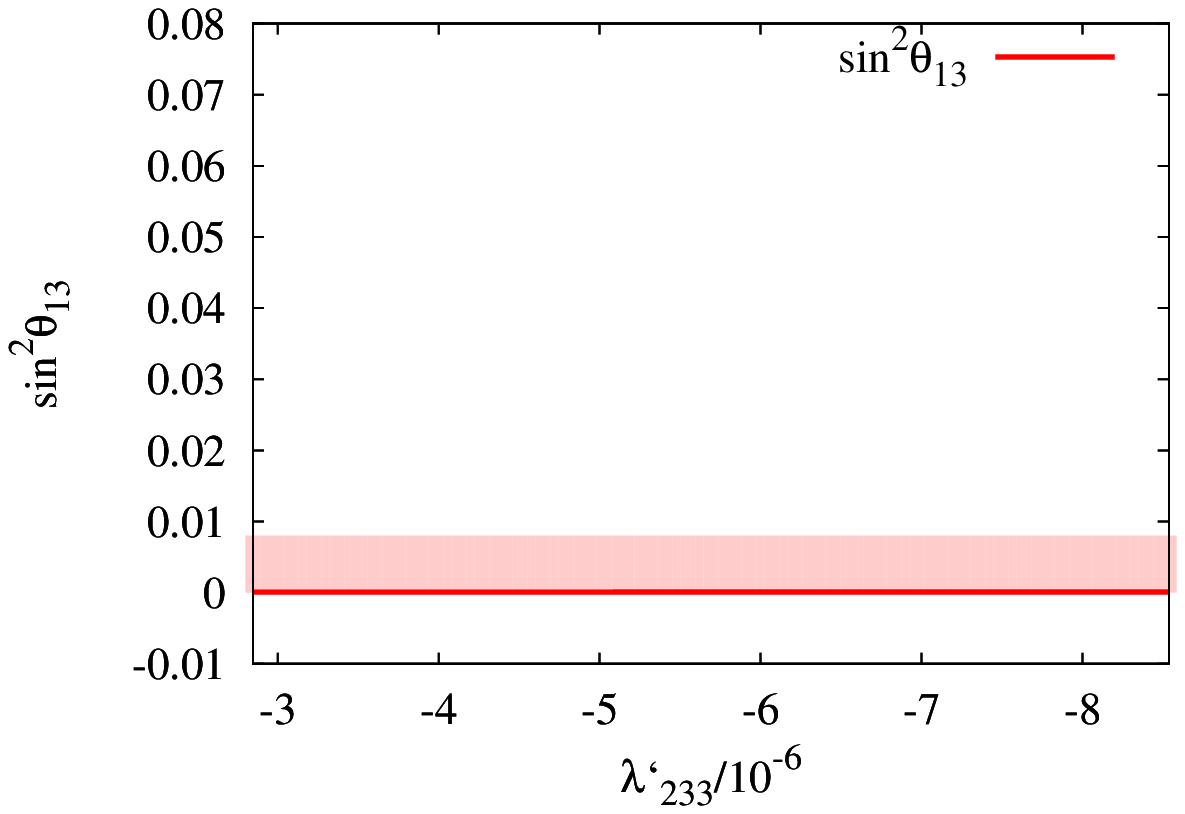} }
\subfigure[$\textrm{sin}^2\theta^{}_{12}$ and $\textrm{sin}^2\theta^{}_{23}$ vs $\lam'_{233}$]{\label{fig:nifull2_lampTheta1223.eps}\includegraphics[scale=0.5]{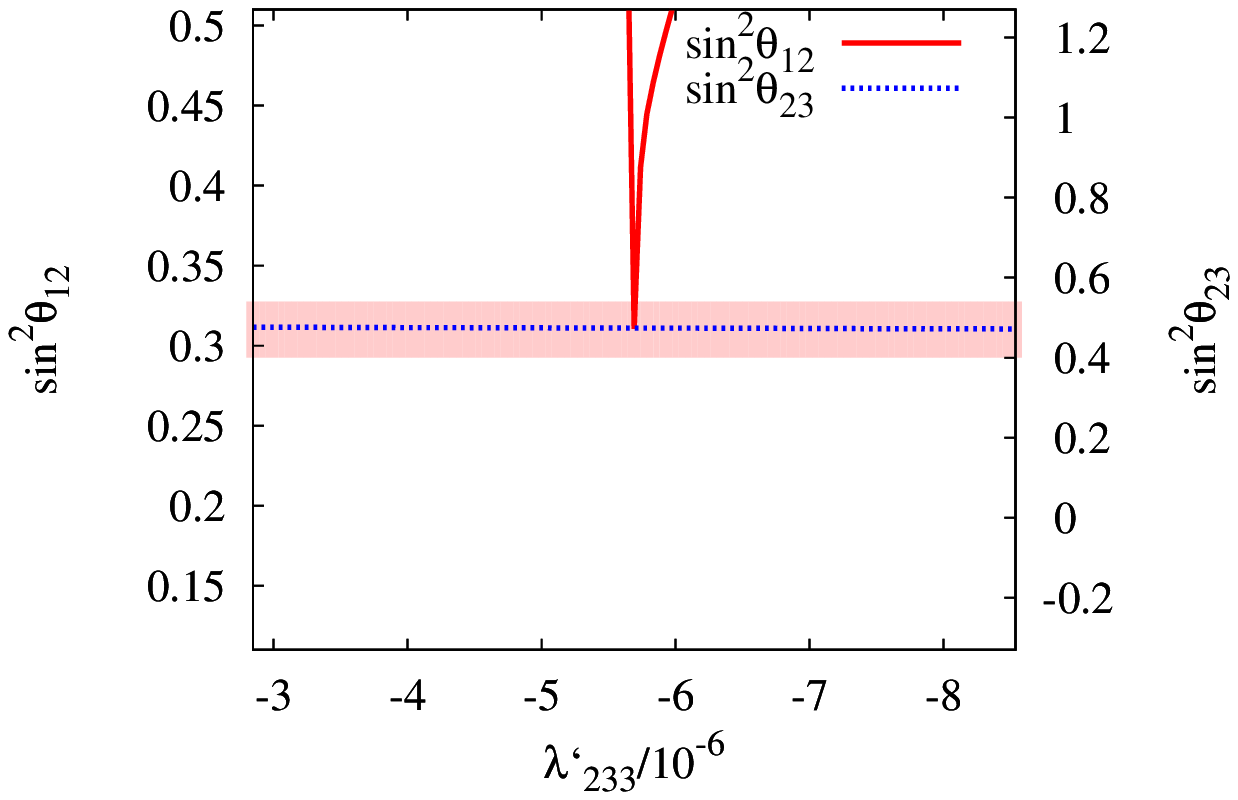} }
\caption{Variation of the observables characterising the neutrino mass pattern with $\lambda'_{233}$ at SPS1a with an inverted hierarchy.  The best fit parameters are $\theta^{l}_{12}=1.55779$, $\theta^{l}_{13}=0.815115$, $\theta^{l}_{23}=0.146126$, $\lam_{233}=1.36023\times 10^{-4}$ and $\lam'_{233}=-5.69116\times 10^{-6}$.  The strips display the empirical $1 \sigma$ bands.  The lines $\theta^{}_{13}$ and $\theta^{}_{23}$ are essentially flat.}\label{fig:invertedhierarchy}
\end{figure}

In Fig.~\ref{fig:invertedhierarchy} we show an example with an inverted
hierarchy, again with non-zero GUT-scale $\lam_{233}$ and $\lam'_{233}$.  The mass
squared differences is much more sensitive to the variation of $\lam'_{233}$,
because of the larger cancellations required to obtain two near degenerate
massive neutrinos.  This reflects that fine tuning between the a priori
unrelated $\lam_{233}$ and $\lam'_{233}$ couplings is required to generate the two
quasi-degenerate mass scales in the inverted hierarchy.  
On the other hand, we
see in Fig.~\ref{fig:nifull2_lampTheta13.eps} and
 Fig.~\ref{fig:nifull2_lampTheta1223.eps} that $\textrm{sin}^2\theta^{}_{13}$ and
$\textrm{sin}^2\theta^{}_{23}$ are insensitive to changes in $\lam'_{233}$.
This is because the lightest neutrino is almost massless, hence its flavour
composition depends only weakly on variations in $\lam'_{233}$ once the mass
degeneracy of the two initially light neutrinos are resolved by the presence
of $\lam'_{233}$.  With an inverted hierarchy, this implies that the third
column of 
$Z_{\nu}$ is insensitive to the changes in the LNV couplings, which in the
standard parameterization is determined fully by $\theta_{13}$ and
$\theta_{23}$\footnote{There is also a dependence on the Dirac CP violating
  phase $\delta$, but it is suppressed by the smallness of
  $\sin^2\theta_{13}$, and we assume CP conservation throughout our numerical
  analysis.}.  Of course, $\textrm{sin}^2\theta^{}_{12}$ will still change as
shown in Fig.~\ref{fig:nifull2_lampTheta1223.eps}, which again is associated
with the cross-over of the two heavy mass eigenstates. 

For an inverted hierarchy, in principle there may be sizeable two loop corrections which could affect the delicate balance among terms in the effective neutrino mass matrix.  However, the two loop corrections in question are not in the literature.  We expect the main corrections to come from the strong interaction, and possibly from heavy flavours.  Smaller corrections may be found, for example, in \cite{0207184}.  We attempted to estimate these 2-loop effects by scaling the 1-loop quark contributions.  For the parameter sets displayed in Tables~\ref{tab:normal result} and~\ref{tab:inverted result}, it was shown that an universal 10\% variation leads to a change in $\chi^2$ of the order 30 for an inverted hierarchy.  However, after re-fitting all $\chi^2$'s go back to the level before scaling, and the change in the best fit parameters is negligible.  For a normal hierarchy, the change in $\chi^2$ is of the order 0.01, and the changes in re-fitted parameters as well as $\chi^2$ are again negligibly small.  Therefore, although more careful analysis of the two loop effects is probably required, we do not expect the possibility of finding modified best fit solutions with two nearly degenerate mass eigenstates in an inverted hierarchy to be affected.

We have described in the previous sections that the tree level mass needs to be suppressed relative to a `natural' value of order given by Eq.~(\ref{eq:tree scale approx}).  As shown in Fig.~\ref{fig:normalhierarchy} and Fig.~\ref{fig:invertedhierarchy}, this can be achieved by a partial cancellation between the two LNV couplings.  To quantify the amount of tuning required, we follow Ref.~\cite{EllisEtAl1986} and define a fine tuning measure $\Delta_{FT}$ by
\eqa \label{eq:fine tune}
\Delta_{FT} &=& |\frac{\partial\textrm{ln}(\mathcal{O(\lam')})}{\partial\textrm{ln}\lam'}|,
\qea
where $\mathcal{O}$ is either $\Delta m^2_{21}$ or $|\Delta m^2_{31}|$, whichever gives a larger $\Delta_{FT}$.  In can be seen in Fig.~\ref{fig:nfull0_lampMM.eps} that for a normal hierarchy, the two observables result in similar $\Delta_{FT}$.  However for the inverted hierarchy, $\Delta_{FT}$ is determined by $\Delta m^2_{21}$, see Fig.~\ref{fig:nifull2_lampMM.eps}.  The second case can be understood from the fact that $\Delta m^2_{21}$ arises from the \emph{difference} between $m_1$ and $m_2$, both of which are of order $\mathcal{O}(\sqrt{|\Delta m^2_{31}|})$.  The variation of $\Delta m^2_{21}$ is then expected to be of order $\mathcal{O}(|\Delta m^2_{31}|)$, the latter of which is significantly larger than the solar mass difference. We obtain numerical values for the fine tuning measure by
\eqa
\Delta_{FT} &\simeq& |\frac{\textrm{ln}(\mathcal{O}(\lam'_a))-\textrm{ln}(\mathcal{O}(\lam'_b))}{\textrm{ln}\lam'_a-\textrm{ln}\lam'_b}|,
\qea
where $\lam'_b$ is the best fit value, and $\lam'_a$ is given by
\eqa
\lam'_a &=& \lam'_b\times(1-2.0\times 10^{-4}).
\qea
For the normal hierarchy, $\Delta_{FT}$ works out to be of order $\mathcal{O}(10)$, so
 the level of fine tuning is mild compared with other SUSY parameters (see e.g. Ref.~\cite{0703044}).
  However for the inverted hierarchy, $\Delta_{FT}$ is typically of order $\mathcal{O}(700-1000)$.
  In this sense, the present study shows a preference towards a normal hierarchy.

In Tables~\ref{tab:normal result} and~\ref{tab:inverted result}, we present a collection of
 parameter sets that minimizes the $\chi^2$ defined in section~\ref{sec:numerics}.  We also
 display the values of $\Delta_{FT}$.  Naturally not all parameter combinations result in good
 fits of the observed parameters listed in Eq.~(\ref{eq:neutrino_data}).  Note however that we do not attempt to exhaust all possibilities
 within this 2 LNV parameters setting.  Also, not all LNV parameter combinations result in
 satisfactory numerical convergence.  While some is due to flavour structure incompatible with
 the requirement of tree level suppression, others are due to problems encountered in the minimization,
 for example running into a corner in the parameter space.  Bearing this in mind, a few
 comments can be made:
\begin{itemize}
\item Because the LNV couplings involved are all very small, the sparticle
  mass spectrum remains essentially the same as in SPS1a, and all branching
  ratios of the sparticles are unaffected except for that of the LSP.
\item It is not possible to obtain the observed mass hierarchy with 2 $\lambda'$'s,
 because in order to suppress the tree-loop mass scale difference, the two $\lambda'$'s
 need to violate the same lepton flavour in order to contribute to the same $\mu_i$
 for the cancellation to occur.  However, with only one flavour violated, there
 can only be one massive neutrino.
\item The successful parameter sets that fit the data all consist of one
  $\lambda$ and one 
  $\lambda'$ coupling.  In principle it is possible to have two $\lambda$'s,
  which 
  might be more interesting because this implies the LSP will decay almost
  exclusively via leptonic channels.  Indeed if the assumption of
  $Z_{lL}=Z_{lR}$ is relaxed, we find parameter sets with two dominant
  $\lambda$ couplings. 
\item Typically all three charged lepton mixing angles are large.  This means
  that we expect many competing decay channels for the LSP, even though we
  only have 2 non-zero LNV couplings in the weak interaction basis.  As an
  example, we display in Tables~\ref{tab:full0_decay} and~\ref{tab:ifull2_decay}
 the major LSP decay channels with $\lambda_{233}$ and
  $\lambda'_{233}$, for a normal hierarchy and for an inverted hierarchy
  respectively. 
\item Although we focus on the SPS1a point, in which the LSP is a neutralino,
  it is also interesting to consider scenarios with e.g. stau LSP.  In the
  latter case, a back of an envelope calculation shows that for a 4 body
  decay to dominate, one typically requires (in the diagonal charged lepton
  basis) LNV couplings with no stau index to be roughly 4 orders of magnitude
  larger than the couplings with a stau index.  This seems unlikely given the
  necessarily large charged lepton mixings. 
\item The lightest neutrino mass is negligible in the current set up.  Given this,
 the other two neutrino masses can be obtained using the atmospheric and solar mass
 squared values.  Regardless of whether one assumes the inverted or natural
 hierarchy, the resulting neutrino mass
 sum automatically satisfies the cosmological bound $\sum_i m_{\nu_i} < 0.7$ eV.  However,
 the size of the couplings at $\MZ$ can still be comparable to the single coupling
 bounds obtained in \cite{0309196} based on the cosmological bound.  This is because
 the required cancellation for the tree level mass suppression makes the couplings
 larger than what would be naively expected.
\item Roughly speaking, for sparticle mass of $\mathcal{O}(100)$ GeV, the LNV coupling products involving $\lam\lam$ or $\lam'\lam'$ need to be of order $\sim\mathcal{O}(10^{-6})$ to have any chance to saturate the $BR(\mu\to e\gamma)$ bound, and are less stringent for the other two leptonic FCNC.  These can be estimated readily using the analytical expressions in Appendix~\ref{llgamma}.  For $BR(B_s \to \mu^+\mu^-)$, the coupling products need to be at least of $\mathcal{O}(10^{-5})$ \cite{9701283}, whereas for $BR(b\to s\gamma)$ the coupling products need to be of $\mathcal{O}(10^{-2})$ \cite{0004067}.  Clearly the couplings involved in our cases are too small to contribute significantly to these rare decays, but we have also checked numerically that this is indeed the case.
\item If the LSP is a stau, it is expected to decay promptly at the interaction point.  If the LSP is a neutralino, it decays into 3 SM fermions.  Assuming a pure photino LSP, an order of magnitude estimate for the 3 body decay is given by \cite{BaerTata}
\eqa
\Gamma(\tilde{\chi}^0_1\to fff)&=&\frac{n_c\alpha(\lambda\textrm{ or }\lambda')^2}{128\pi^2}\frac{m_{\tilde{\chi}^0_1}^5}{M^4_{\textrm{SUSY}}},
\qea
where the $f$'s denote SM fermions, $n_c$ is the number of colours, $\alpha$ is the fine structure constant, and $M_{SUSY}$ is the mass scale of the virtual, intermediate scalar.  For the results displayed in Tables~\ref{tab:normal result} and~\ref{tab:inverted result}, the neutralino can decay with a displaced vertex of $\mathcal{O}(0.1)$mm (the total width is of order ${\mathcal O}(10^{-12}-10^{-11})$GeV).  Such a displacement would not be immediately obvious, but could be searched for, providing additional confirmation.

\item For the inverted hierarchy, due to the high degree of tuning, all digits
  presented in Table~\ref{tab:inverted result} are needed in order to
 produce the correct neutrino
  oscillation parameters. 
\end{itemize}

\TABLE
{
\scalebox{0.94}{
\small
  \begin{tabular}{|ll|lll|l|}
    \hline
    $\Lambda_1$&$\Lambda_2$&$\theta^l_{12}$&$\theta^l_{13}$&$\theta^l_{23}$&$\Delta_{FT}$ \\
    \hline
    $^{a}\lambda'_{233}=-2.49978\times 10^{-6}$ &$\lambda_{233}=4.06508\times 10^{-5}$& 0.459520& 0.388989& 0.304863 & 8.09\\
    $^{a}\lambda'_{233}=-2.50019\times 10^{-6}$ &$\lambda_{211}=4.06533\times 10^{-5}$& 1.98935& 1.08162& 0.632130 & 8.10\\
    $^{a}\lambda'_{233}=-3.41336\times 10^{-6}$ &$\lambda_{321}=9.86746\times 10^{-5}$& 0.448321& 0.400030& 2.89062 & 12.4\\ \hline
    $^{b}\lambda'_{122}=-1.14066\times 10^{-4}$ &$\lambda_{122}=4.06346\times 10^{-5}$& 1.19298& 0.190538& 1.17391 & 11.8\\
    $^{b}\lambda'_{122}=-8.97777\times 10^{-5}$ &$\lambda_{123}=1.02771\times 10^{-4}$& 2.10672& 0.174800& 1.18124 & 9.44\\
    $^{b}\lambda'_{122}=-8.59626\times 10^{-5}$ &$\lambda_{133}=4.09647\times 10^{-5}$& 0.997963& 0.281922& 0.417935 & 8.00\\\hline
    $^\dagger\lambda'_{311}=-8.34433\times 10^{-4}$ &$\lambda_{311}=4.09870\times 10^{-5}$& 2.01028& 1.07090& 2.18984 & 8.00\\
    $^\dagger\lambda'_{311}=-1.14183\times 10^{-3}$ &$\lambda_{321}=9.88004\times 10^{-5}$& 0.448302& 0.400044& 1.32004 & 12.4\\
    $^\dagger\lambda'_{311}=-8.66312\times 10^{-4}$ &$\lambda_{132}=9.85792\times 10^{-5}$& 0.921694& -1.12754& 2.00207 & 9.26\\\hline
    $^\dagger\lambda_{233}=-4.07409\times 10^{-5}$ &$\lambda'_{211}=8.36075\times 10^{-4}$& 0.457859& 0.390978& 0.304896 & 8.08\\
    $^\dagger\lambda_{233}=-4.07226\times 10^{-5}$ &$\lambda'_{221}=4.16596\times 10^{-4}$& 0.458736& 0.389932& 0.304588 & 8.09\\
    $^\dagger\lambda_{233}=-4.06542\times 10^{-5}$ &$\lambda'_{231}=7.40962\times 10^{-4}$& 0.459528& 0.388981& 0.304805 & 8.09\\\hline
  \end{tabular}
}
  \caption{A selection of LNV parameters and charged lepton mixings at SPS1a
  which fit the neutrino masses and mixings for a normal hierarchy.
  All the points shown have $\chi^2 < 10^{-3}$.  The parameter sets
  marked with $\dagger$ are ruled out by the $\mu$Ti$\to$ $e$Ti constraints in
  \cite{9906209}.} \label{tab:normal result}
}

\TABLE
{
\scalebox{0.94}{
\small
  \begin{tabular}{|ll|lll|l|l|}
    \hline
    $\Lambda_1$&$\Lambda_2$&$\theta^l_{12}$&$\theta^l_{13}$&$\theta^l_{23}$&$\Delta_{FT}$&$\chi^2$ \\
    \hline
    $^{a}\lambda'_{233}=-5.69116\times 10^{-6}$ &$\lambda_{233}=1.36023\times 10^{-4}$& 1.55779& 0.815115& 0.146126& 755& 0.01\\
    $^{a}\lambda'_{233}=-5.69126\times 10^{-6}$ &$\lambda_{211}=1.32365\times 10^{-4}$& 1.38843& 0.760045& 0.140903& 758& 0.06\\
    $^{a}\lambda'_{233}=-5.67940\times 10^{-6}$ &$\lambda_{123}=1.42938\times 10^{-4}$& 1.81386& -0.757538& 0.141975& 726&
    0.05\\ \hline
    $^{b}\lambda'_{122}=-1.96283\times 10^{-4}$ &$\lambda_{122}=1.24824\times 10^{-4}$& 0.134765& 0.101938& 0.798282& 988& 0.43\\
    $^{b}\lambda'_{122}=-1.93673\times 10^{-4}$ &$\lambda_{132}=1.47364\times 10^{-4}$& 3.03234& 0.0866645& 0.931616& 743& 2.85\\
    $^{b}\lambda'_{122}=-1.96175\times 10^{-4}$ &$\lambda_{123}=1.45708\times 10^{-4}$& 0.144386& 0.0943266& 0.689708& 736& 0.52\\\hline
    $^\dagger\lambda'_{311}=-1.90386\times 10^{-3}$ &$\lambda_{311}=1.16446\times 10^{-4}$& 1.76438& 0.833990& 1.40752& 767& 1.41\\
    $^\dagger\lambda'_{311}=-1.90027\times 10^{-3}$ &$\lambda_{321}=1.44392\times 10^{-4}$& 1.52379& 0.854066& 1.42182& 719& 0.22\\
    $^\dagger\lambda'_{311}=-1.90809\times 10^{-3}$ &$\lambda_{322}=1.37574\times 10^{-4}$& 1.57829& 0.820421& 1.42557& 984& 0.01\\\hline
    $^\dagger\lambda_{233}=-1.36245\times 10^{-4}$ &$\lambda'_{211}=1.90793\times 10^{-3}$& 1.55054& 0.815304& 0.146304& 979& 0.01\\
    $^\dagger\lambda_{233}=-1.36235\times 10^{-4}$ &$\lambda'_{221}=9.50337\times 10^{-4}$& 1.55302& 0.815293& 0.146092& 975& 0.00\\
    $^\dagger\lambda_{233}=-1.36016\times 10^{-4}$ &$\lambda'_{231}=1.68685\times 10^{-3}$& 1.56378& 0.815077& 0.145634& 753& 0.00\\\hline
  \end{tabular}
}
  \caption{A selection of LNV parameters and charged lepton mixings at SPS1a
  which fit the neutrino masses and mixings for an inverted hierarchy.
  The parameter sets marked with $\dagger$ are ruled out by the $\mu$Ti $\to$
  $e$Ti constraints in \cite{9906209}.  Note that because of the high fine
  tuning, all digits present in the table are required to obtain the correct numerical
  results.} 
  \label{tab:inverted result} 
}

As mentioned in section~\ref{sec:numerics}, the numerical analysis is
performed in a basis where all quark mixings reside in the down quark sector.
In the case of up type quark mixings, only $\lam'$ couplings with the same
quark flavour indices can be used to balance the contribution of $\lam$ in the
dynamical generation of the bilinear parameters, because the down quark Yukawa
matrix $Y_D$ is diagonal in this basis.  Apart from this, the main effect in
changing the quark mixing is expected to be a rescaling of $\lam'$, with only
minor changes to $\lam$ and the charged lepton mixing angles.  This is
confirmed by our numerical results. 

In Tables~\ref{tab:full0_decay} and~\ref{tab:ifull2_decay}, the branching
ratios of the neutralino LSP for non-zero $\lambda_{233}$ and $\lambda'_{233}$
with the normal and inverted hierarchy respectively are generated using
\isajet \cite{isajet}.  In both cases there are many competing channels for
the LSP decay, not surprising due to the typically large charged lepton
mixing.  On the other hand, it is amusing to observe the approximate left
handed $\mu-\tau$ symmetry for the normal hierarchy, and approximate
$\mu-\tau$ symmetry for \emph{both} chiralities in the inverted hierarchy.
This corresponds to an approximate symmetry among the $\lambda$ couplings after
rotating to the diagonal charged lepton basis, and should be compared with the
approximate $\mu-\tau$ symmetry in the experimentally favored near tri-bi
maximal neutrino mass matrix \cite{0707.2713}.  This effect appears in all
parameter sets we have checked which give rise to the observed neutrino mass
pattern.  However the result depends on the assumed condition of
$Z_{lL}=Z_{lR}$.  We have checked that in other basis, for example with
$Z_{lR}=1_{3\times 3}$, this approximate symmetry ceases to appear. 

\DOUBLETABLE{
\scalebox{0.92}{
  \begin{tabular}{|l|l|l|l|l|}
    \hline
    $\lam_{ijk}$& Channel & BR & Channel & BR \\
    \hline
    $\lam_{122}$& $e^-$ $\numu$ $\mu^+$ & 0.006 & $\mu^-$ $\nue$ $\mu^+$ & 0.006\\
    $\lam_{132}$& $e^-$ $\nutau$ $\mu^+$ & 0.007 & $\tau^-$ $\nue$ $\mu^+$ & 0.007\\
    \hline
    $\lam_{123}$& $e^-$ $\numu$ $\tau^+$ & 0.029 & $\mu^-$ $\nue$ $\tau^+$ & 0.029\\
    $\lam_{133}$& $e^-$ $\nutau$ $\tau^+$ & 0.034 & $\tau^-$ $\nue$ $\tau^+$ & 0.034\\
    \hline
    $\lam_{231}$& $\mu^-$ $\nutau$ $e^+$ & 0.005 & $\tau^-$ $\numu$ $e^+$ & 0.005\\
    \hline
    $\lam_{232}$& $\mu^-$ $\nutau$ $\mu^+$ & 0.027 & $\tau^-$ $\numu$ $\mu^+$ & 0.028\\
    \hline
    $\lam_{233}$& $\mu^-$ $\nutau$ $\tau^+$ & 0.138 & $\tau^-$ $\numu$ $\tau^+$ & 0.140\\
    \hline
  \end{tabular}
}}
{
\scalebox{0.92}{
  \begin{tabular}{|l|l|l|l|l|}
    \hline
    $\lam_{ijk}$ & Channel & BR & Channel & BR\\
    \hline
    $\lam_{122}$ & $e^-$ $\numu$ $\mu^+$  & 0.063 & $\mu^-$ $\nue$ $\mu^+$ & 0.063\\
    $\lam_{132}$ & $e^-$ $\nutau$ $\mu^+$ & 0.056 & $\tau^-$ $\nue$ $\mu^+$ & 0.057\\
    \hline
    $\lam_{123}$ & $e^-$ $\numu$ $\tau^+$ & 0.067 & $\mu^-$ $\nue$ $\tau^+$ & 0.067\\
    $\lam_{133}$ & $e^-$ $\nutau$ $\tau^+$& 0.059 & $\tau^-$ $\nue$ $\tau^+$ & 0.060\\
    \hline
  \end{tabular}
}}
{Decay channels of the LSP $\neut_1$ at SPS1a, with 
  $\lam_{233}$ and $\lam'_{233}$ at $\MX$ given in the top row of
 Table~\protect\ref{tab:normal result}, and a normal hierarchy.  Each process has a charged conjugated counterpart which is not shown in the table.  Only branching ratios $\geq$ 0.005 are displayed.\protect\label{tab:full0_decay}}
{Decay channels of the LSP $\neut_1$ at SPS1a, with $\lam_{233}$ and
  $\lam'_{233}$ at $\MX$ shown in the top row of Table~\protect\ref{tab:inverted result}, and an inverted hierarchy.  Each process has a charge-conjugated counterpart which is not shown in the table.  Only branching ratios $\geq$ 0.005 are displayed.\protect\label{tab:ifull2_decay}}

Another observation is that the dominant decay channels in these two samples
are purely leptonic despite the presence of the $\lam'$ couplings.  This is
mainly due to the choice of the third family indices in $\lambda'_{233}$, which
involves bottom Yukawa couplings in the dynamical generation of $\mu_i$.  The
cancellation condition then requires this coupling to be relatively small.  In
cases with $\lambda'$ coupling which involves smaller quark Yukawa couplings,
the 3 body decay via this coupling can become dominant. Then, LSP decays into
a charged lepton plus jets and some SUSY production events may be fully
reconstructed since there would be no intrinsic missing energy.

\section{Discussion}\label{sec:conclusion}

We have shown that it is possible to obtain the observed neutrino mass pattern
with 2 trilinear lepton number violating parameters in the mSUGRA set up in a
basis where the lepton Yukawa coupling matrix is non-diagonal.  However,
depending on the 
flavour structure, the two non-zero couplings can either have one dominate over
another, or have comparable magnitude. Here, we do not address the origin of
hierarchies in the GUT-scale dimensionless couplings. They may be set, for
example, by a spontaneously broken U(1) family gauge
symmetry~\cite{Rossandibanez}. 
As an illustration, the first three
parameter sets shown, labelled with superscripts (a), in both Tables~\ref{tab:normal result} and~\ref{tab:inverted result} have $\lam'_{233}$ a factor
$\sim {\mathcal O}(10)$ smaller than the other LNV operator and
so may be described by flavour models with only one dominant operator.  For the
next three parameter sets labelled with superscripts (b), $\lam_{1ij}$ and
$\lam'_{122}$ are of similar magnitude.  A rough indicator of whether
particular LNV parameters may contribute significantly to the neutrino mass
may be defined by noting that the dynamical generation of $\mu_i$ scales like
$\lam (Y_E^*) + 3 \lam' (Y_D^*)$ from Eq.~(\ref{eq:dyn gen}).  As an example
suppose we are able to identify one coupling, say $\lam_{ikl}$, that
contribute to the neutrino mass matrix significantly.  Other couplings will
also be important if  
\eqa
\frac{\lam_{ikl}}{\lam'_{jmn}} &\simeq& \frac{(Y_D^*)_{mn}}{(Y_E^*)_{kl}},
\qea
where the Yukawa couplings are evaluated in the weak interaction basis.  Whether we have single coupling dominance or not can then be estimated by the ratio of the Yukawa couplings.  However, for an inverted hierarchy, a slight change in the neutrino mass matrix can modify the results significantly, so more care is needed when applying this rough indicator.

In the following we briefly compare our models with the literature.  In neutrino models with bilinear R-parity violation, e.g. \cite{9901254, 9511288, 9606388}, the vacuum oscillation solution is generally favoured.  This is due to the alignment effect which creates a large hierarchy between the tree level mass and the loop induced masses.  Because we are interested in models with only trilinear R-parity violation at the unification scale, we make no further comments on this and refer interested readers to the literature.

In \cite{9903435}, models with many $\lambda_{ijk}'$ couplings are discussed.
It was shown how such models naturally lead to the now excluded vacuum
oscillation solution of the solar neutrino problem.  Our work complement such
results by discussing the issues need to be addressed for a near tri-bi
maximal mixing solution, and also performs the numerical calculation by
considering the full set of RGEs and include all one-loop contributions to the
neutrino masses. 

In \cite{9807327}, fits on the neutrino masses in LNV mSUGRA using trilinear
couplings in the diagonal lepton basis were attempted.  The need
for tree level mass suppression and the need to break
the alignment effect were also
discussed there.  
A hierarchy was assumed which included only the terms $\lam'_{i33}$ and/or $\lam_{i33}$ as a first approximation.  
This requires a different tree mass suppression method involving the RPC soft SUSY breaking parameters, whereas in our work, the suppression is obtained through interplay among the LNV parameters.  The RGEs used in our work also included the contributions of the quarks and leptons from the first and second generations to the soft SUSY breaking terms.  
These can be important as, depending on their size, the LNV couplings with
first and second generation indices may contribute significantly to the
neutrino masses through coupling to the RPC Yukawa matrices in the RG
equations. 
This also allows us to go beyond the natural hierachy assumed in
\cite{9807327}.  Preference towards the small mixing angle MSW solution was
preferred in their analysis, but only the mass squared differences and the
atmospheric mixing angle was fitted.  In our work, we obtain the best fit
values of the LNV parameters with all 3 PMNS mixing angles as well as the mass
squared differences. 

In \cite{0603225}, the authors described how the tri-bi maximal mixing pattern
may be obtained by setting particular combinations of LNV parameters at the
electroweak scale, which essentially involves setting two set of parameters,
each with a different scale, to fit the observed neutrino mass pattern with
the two scales $|\Delta m^2_{31}|$ and $\Delta m^2_{21}$.  Besides working in
different basis, our models are in general inequivalent, because by setting
the LNV couplings at $\MZ$, the problem associated with a tree-loop hierarchy
is absent, and the mass scales involved can be set directly, without any need
of tree-loop level cancellation.  Depending on the parameters chosen, it is
even possible to have a vanishing tree level contribution in such models.  The
models presented in this paper are more difficult to control in this sense
because the two mass scales involved are related by the (suppressed) dynamical
generation of the bilinear terms. 
In a similar spirit, Ref.~\cite{0711.4315} assumes
ansatzes for the weak-scale tri-linear LNV couplings in terms
of six parameters. Some resulting models were shown to fit low-energy bounds as
well as the atmospheric and solar neutrino anomalies. Some
implications for LHC signals were investigated.

Finally, we comment on the amount of tuning required.  It may appear that the
cancellation required is unnatural.  To some
extent we do need certain degree of tuning.
We must cancel the tree-level mass sufficiently such that
the loop corrections are roughly $\mathcal{O}(0.1)$ times the tree level mass
for a normal hierarchy.  This implies that $\mu_i$ is suppressed by
$\mathcal{O}(1)$ compared to its `natural' scale.  This intuition is in
agreement with the fine tuning measure discussed in section~\ref{sec:results}.
  For an inverted hierarchy, the loop corrections need to be
roughly of the same order of magnitude as the tree level piece, and the solar
neutrino mass difference arises from the difference between the masses of the
two heaviest 
neutrinos, both of which are of the atmospheric mass scale.  This means that
while the parameters do need to be fine tuned for an inverted hierarchy due to
the small solar neutrino mass difference compared to the atmospheric mass
difference, for the normal hierarchy the tuning is reasonably mild.  Once the
tree and the loop contributions to the neutrino masses become comparable, we
see that the suppression of tree level mass also leads to departure from
alignment, and significant neutrino mixing results if one allows for charged
lepton mixings, possibly provided by some underlying flavour model.  Thus
large PMNS mixing is not unreasonable, although we have to treat the
experimentally favoured near tri-bi maximal neutrino mass pattern as accidental
in our approach. 

To summarise, we have investigated the minimal configuration in mSUGRA with
trilinear R-parity violation that can reproduce the observed neutrino
oscillation observables.  We obtain this by adding lepton \emph{number}
violating parameters in a basis where the charged lepton Yukawa matrix is not
diagonal.  We discuss on general grounds that it is impossible to obtain the
neutrino mass hierarchy with just one LNV parameter in such a basis. This can
be traced to 
the fact that the bilinear lepton number violating operators responsible for
the tree level mass generation are radiatively generated by the trilinear
coupling.  Despite the radiative origin of the bilinear operators, it turns
out the tree level mass scale is dominant over the loop mass scale.  Another
general result is the alignment effect, whereby the tree level mass matrix and
the loop corrections have almost the same pattern, therefore suppressing the
light neutrino mass further. 

These effects can already be alleviated by introducing a second trilinear
operator at $\MX$.  The reason is that the bilinear operators will be
suppressed if the trilinear operators contribute to their dynamical
generations in an opposite way.  
The introduction of another dominant coupling also weakens the alignment,
 and the PMNS mixings observed can be obtained with large charged lepton mixings. 

We present, for both hierarchies, several data sets at SPS1a obtained by
minimizing the chi-square of the neutrino oscillation observables.  The
couplings involved are too small to affect the spectrum and decay of the
sparticles, except for the LSP, which has to decay via the LNV operators.  We
find typically that there are a number of significant decay channels, while
the actual channels are model dependent.  Depending on the choice of the
$\lam'$ coupling, certain parameter sets are ruled out by the $\mu-e$
conversion in nuclei, but the LNV contributions to the branching ratios for
other 
FCNCs considered are typically not experimentally accessible. 

While we need a certain degree of tuning to ensure the mass ratios fall in the
observed range, the degree of tuning required for a normal mass hierarchy is
relatively mild with $\Delta_{FT}\sim \mathcal{O}(10)$.  The fine tuning required
 for the inverted hierarchy $\Delta_{FT}\sim \mathcal{O}(800)$ reflects
the need to arrange two a priori unrelated parameters to give two (quasi-)
degenerate masses to account for the small solar neutrino mass squared difference.

Evidently with so many lepton number violating operators, there are many ways
to reconstruct the neutrino mass matrix.  While we have 
shown
that it is possible to build simple models with a manageable LNV parameter
space, future work is required to realize the many other possibilities within
this class of models.

\section*{Acknowledgement}
This work has been partially supported by STFC. 
We thank the members of the Cambridge SUSY working group and S Rimmer for
useful conversations.  The computational work has been performed using the
Cambridge 
eScience CAMGRID computing facility, with the invaluable help of Mark Calleja.
CHK is funded by a Hutchison Whampoa DHPA studentship. 

\vspace*{2.0cm}
\appendix
\section{Mass Matrices} \label{Appendix: mass matrices}
In this section we write down the CP-even, CP-odd, neutral fermion and charged
fermion mass matrices.  The notation follows that in \cite{0309196}, where the
mass matrices are presented in terms of couplings in the Lagrangian.  However
following \softsusy, the Standard Model Higgs vev is normalized to $v \approx
246\textrm{GeV}$.  The `t Hooft Feynman gauge is used for numerical
computation, although we present the mass matrices in the general $\xi$
gauge. 

Our convention is the same as in \cite{0603225} where all mass matrices are
presented in 2-component Weyl notation (apart from the inclusion of the
sneutrino vevs).  We also follow their definition of the mixing matrices.  In
the R-Parity conserving case, the mass matrices reduce to those in
\cite{9606211}\footnote{In Ref.~\protect\cite{9606211} the mass matrices are
  presented  in 4-component 
  Dirac notation.}. 

With LNV, the down type higgs and the lepton doublets are put on the same
footing.  It is therefore instructive to write down the superpotential and the
Lagrangian in a way that express this equivalence.  We define 
\eqa
\mathcal{L}_{\alpha} &\equiv& (H_d, L_i),
\qea
where $\alpha, \beta, \ldots \in \{ 0,1,2,3 \}$ and $i,j,k,\ldots \in \{ 1,2,3
\}$. 
The parameters are defined by
\eq
\begin{array}{rclrcl}
\lam_{\alpha\beta k} &\equiv& \Big(\lam_{0jk}\equiv(Y_E)_{jk},\lam_{ijk}\Big),\qquad & h_{\alpha\beta k} &\equiv& \Big(h_{0jk}\equiv(h_E)_{jk}, h_{ijk}\Big),\\
\lam'_{\alpha jk} &\equiv& \Big(\lam'_{0jk}\equiv(Y_D)_{jk}, \lam'_{ijk}\Big), & h'_{\alpha jk} &\equiv& \Big(h'_{0jk}\equiv(h_D)_{jk}, h'_{ijk}\Big), \\
\mu_{\alpha} &\equiv& \Big(\mu, \mu_i\Big), & b_{\alpha} &\equiv& \Big(\tilde{B}, \tilde{D}_i\Big),
\end{array}
\qe
and
\eqa
(m^2_{\mathcal{L}})_{\alpha\beta} \equiv \Big((m^2_{\mathcal{L}})_{00}\equiv m^2_{h_d}, (m^2_{\mathcal{L}})_{0i}\equiv m^2_{h_dL_i}, (m^2_{\mathcal{L}})_{ij}\equiv (m^2_{\tilde{L}})_{ij}\Big).
\qea
In this notation, the superpotential is written as
\eqa
\mathcal{W}&=&\frac{1}{2}\lam_{\alpha\beta k}\lag_{\alpha}\lag_{\beta}\bar{E_k} +\lam'_{\alpha jk}\lag_{\alpha}Q_j\bar{D_k} +(Y_U)_{ij}Q_iH_u\bar{U_j} -\mu_{\alpha}\lag_{\alpha}H_u,
\qea
while the soft supersymmetry breaking Lagrangian is denoted to be
\eqa
-\mathcal{L}_{soft}&=&\frac{1}{2}h_{\alpha\beta k}\tilde{\lag}_{\alpha}\tilde{\lag}_{\beta}\sse_k +h'_{\alpha jk}\tilde{\lag}_{\alpha}\ssq_j\ssdown_k  -b_{\alpha}\tilde{\lag}_{\alpha}h_u +(h_U)_{ij}\ssq_ih_u\ssup_j + h.c. \nonumber \\
&& +\tilde{\lag}^{\dagger}(m^2_{\tilde{\lag}})\tilde{\lag} +\sse^{\dagger}(m^2_{\sse})\sse +m^2_{h_u}h_u^{\dagger}h_u \nonumber +\ssq^{\dagger}(m^2_{\ssq})\ssq  +\ssup^{\dagger}(m^2_{\ssup})\ssup  +\ssdown^{\dagger}(m^2_{\ssdown})\ssdown \nonumber \\
&& +
\left[ \frac{1}{2}M_1\tilde{B}\tilde{B} +\frac{1}{2}M_2\tilde{W}\tilde{W}
  +\frac{1}{2}M_3\tilde{g}\tilde{g} +h.c. \right].
\qea
The mass matrices presented below follow the above notation.

\subsection*{Higgs-sneutrino masses}
After electroweak symmetry breaking, the sneutrinos, $\ssnu_i$ mixes with the Higgs bosons $h_u$ and $h_d\equiv\ssnu_0$, resulting in CP-even (CPE) and CP-odd (CPO) scalars.  The CPE and CPO Higgs-sneutrino mass eigenstates are obtained in a generic basis with $\tilde{\nu}_{\alpha}\equiv (h_d,\tilde{\nu}_i)$ after the diagonalization of the $5\times5$ mass matrices
\eqa
\lag &=& -\frac{1}{2}\Big(\textrm{Re}h_u \textrm{ Re}\ssnu_{\gamma} \Big)\mathcal{M}^2_{CPE} \left(\begin{array}{c} \textrm{Re}h_u \\ \textrm{Re}\ssnu_{\delta}\end{array} \right), \\
\lag &=& -\frac{1}{2}\Big(\textrm{Im}h_u \textrm{ Im}\ssnu_{\gamma} \Big)\mathcal{M}^2_{CPO} \left(\begin{array}{c} \textrm{Im}h_u \\ \textrm{Im}\ssnu_{\delta}\end{array} \right),
\qea
where
\eqa
\mathcal{M}^2_{CPE} &=& \left( \begin{array}{cc}
\frac{b_{\alpha}v_{\alpha}}{v_u} & -b_{\delta} \\
 -b_{\gamma} & (m^2_{\tilde{\nu}})_{\gamma\delta} \end{array} \right) 
+ \frac{g^2+g^2_2}{4}\left( \begin{array}{cc} 
v^2_{u} & -v_{u} v_{\delta} \\
-v_u v_{\gamma} & v_{\gamma} v_{\delta} \end{array} \right), \\
\mathcal{M}^2_{CPO} &=& \left( \begin{array}{cc}
\frac{b_{\alpha}v_{\alpha}}{v_u} & b_{\delta} \\
 b_{\gamma} & (m^2_{\tilde{\nu}})_{\gamma\delta} \end{array} \right) 
+ \frac{g^2+g^2_2}{4}\xi\left( \begin{array}{cc} 
v^2_{u} & -v_{u} v_{\delta} \\
-v_u v_{\gamma} & v_{\gamma} v_{\delta} \end{array} \right),
\qea
where $\xi$ is a gauge parameter\footnote{$\xi=1$ corresponds to the `t Hooft Feynman gauge.} and
\eqa
(m^2_{\tilde{\nu}})_{\gamma\delta} &\equiv& [(m^2_{\tilde{\mathcal{L}}})_{\gamma\delta}
  +\mu^*_{\gamma}\mu_{\delta}]-\frac{(g^2+g^2_2)}{8}(v_u^2-|v_{\alpha}|^2)\delta_{\gamma\delta}.
\qea
Here we have $|v_{\alpha}|^2 = v_d^2 + v_i^2$.  The rotation matrices are defined by
\eqa
Z^T_{R}\mathcal{M}^2_{CPE}Z_{R} &=& \textrm{diag}[m^2_{h^0}, m^2_{H^0}, m^2_{\tilde{\nu}^i_+}], \qquad i=1,2,3, \\
Z^T_{A}\mathcal{M}^2_{CPO}Z_{A} &=& \textrm{diag}[m^2_{G^0}, m^2_{A^0}, m^2_{\tilde{\nu}^i_-}], \qquad i=1,2,3.
\qea

\subsection*{Neutrino-neutralino masses}
In a general flavour basis, with $\nu_{\beta}\equiv (\tilde{h}^0_d, \nu_i)$, the $7\times7$ neutrino-neutralino mass matrix is given by
\eqa
\lag &=& -\frac{1}{2}\Big(-i\widetilde{\mathcal{B}}\quad -i\widetilde{\mathcal{W}}^3\quad \tilde{h}_u^0\quad \nu_{\alpha} \Big)\mathcal{M}_N \left(\begin{array}{c}-i\widetilde{\mathcal{B}}\\ -i\widetilde{\mathcal{W}}^3\\ \tilde{h}_u^0\\ \nu_{\beta} \end{array} \right),
\qea
where (at tree level)
\eqa
\mathcal{M}_N &=& \left( \begin{array}{cccc}
M_1 & 0 & \frac{1}{2}gv_u & -\frac{1}{2}gv_{\beta} \\
0 & M_2 &-\frac{1}{2}g_2v_u & \frac{1}{2}g_2v_{\beta} \\
\frac{1}{2}gv_u & \frac{1}{2}g_2v_u & 0 &-\mu_{\beta} \\
-\frac{1}{2}gv_{\alpha} & \frac{1}{2}g_2v_{\alpha} & -\mu_{\alpha} & 0_{\alpha\beta}
\end{array} \right).
\qea
The rotation matrix is given by
\eqa
Z^T_N\mathcal{M}_N Z_N &=& \textrm{diag}[\tilde{\chi}^0_a, \nu_i], \qquad a=1,2,3,4. \quad i=1,2,3.
\qea

\subsection*{Charged lepton-chargino masses}
The charged lepton-chargino mixing is displayed in the $5\times5$ matrix given below
\eqa
\lag &=& -\Big(-i\widetilde{\mathcal{W}}^-\quad e_{L_{\alpha}} \Big)\mathcal{M}_C \left(\begin{array}{c}-i\widetilde{\mathcal{W}}^+\\ \tilde{h}_u^+\\ e^+_{R_k} \end{array} \right) + h.c.,
\qea
where $e_{L_{\alpha}} = (\tilde{h}^-_d, e_L^-)$, and at tree level
\eqa
\mathcal{M}_C &=& \left( \begin{array}{ccc}
M_2 & \frac{1}{\sqrt{2}}g_2v_u & 0_k \\
\frac{1}{\sqrt{2}}g_2v_{\alpha} & \mu_{\alpha} & \frac{1}{\sqrt{2}}\lambda_{\alpha\beta k}v_{\beta}
\end{array} \right).
\qea
The diagonalisation matrices are defined by
\eqa
Z^{\dagger}_-\mathcal{M}_C Z_+ &=& \textrm{diag}[\tilde{\chi}^{\pm}_a, e_i], \qquad a=1,2. \quad i=1,2,3.
\qea


\section{One loop self energies of the neutral fermions}\label{Appendix: self energies}
For completeness, we list in this appendix the general form of the one loop contributions to the neutral fermion.  They can be separated into 4 classes, depending on whether the chirality flip is in the loop or on the external legs, and whether the boson in the loop is a scalar or a vector boson.  The complete listing in the RPC limit including the Feynman rules is in \cite{9606211}.  The Feynman rules for the general LNV calculation can be found in \cite{0603225}.  Our expressions are equivalent to the latter, but for the gauge boson contributions, we use the Passarino-Veltman (PV) `B' functions \cite{PV} only, and present in a form easier to be compared with \cite{9606211}, which adopts the `t Hooft-Feynman gauge ($\Chi=1$).

Recall in the main text that the one loop self energy correction $\delta \mM_{ij}$ takes the form:
\eqa
\delta \mM_{ij} &=& (\Sigma_D)_{ij} - (\mM_N)_{ik}(\Sigma_L)_{kj},
\qea
where $\Sigma_D$ and $\Sigma_L$ are the mass and kinetic term corrections respectively.

\begin{figure}[!ht]
  \begin{center}
    \scalebox{1.1}{
    \subfigure[$\Sigma^{\phi}_D$ contribution]{
    \begin{picture}(140,70)(0,0)
      \DashArrowArcn(70,40)(20,180,360){5}
      \ArrowArcn(70,40)(20,270,180)
      \ArrowArc(70,40)(20,270,360)
      \ArrowLine(20,40)(50,40)
      \ArrowLine(120,40)(90,40)
      \Vertex(70,20){1.5}
      \put(65,25){$m_{\psi}$}
      \put(10,39){$\tilde{\chi}^0_i$}
      \put(122,39){$\tilde{\chi}^0_j$}
      \put(28,25){$a_{\tilde{\chi}^0_i\phi\psi}$}
      \put(95,25){$b_{\tilde{\chi}^0_j\phi\psi}$}
      \put(70,65){$\phi$}
      \put(70,10){$\psi$}
    \end{picture}
    }
    \subfigure[$\Sigma^{\phi}_L$ contribution]{
    \begin{picture}(140,70)(0,0)
      \DashArrowArcn(70,40)(20,180,360){5}
      \ArrowArc(70,40)(20,180,0)
      \ArrowLine(15,40)(32.5,40)
      \ArrowLine(50,40)(32.5,40)
      \ArrowLine(120,40)(90,40)
      \Vertex(32.5,40){1.5}
      \put(5,39){$\tilde{\chi}^0_i$}
      \put(122,39){$\tilde{\chi}^0_j$}
      \put(30,25){$b^*_{\tilde{\chi}^0_k\phi\psi}$}
      \put(95,25){$b_{\tilde{\chi}^0_j\phi\psi}$}
      \put(70,65){$\phi$}
      \put(70,10){$\psi$}
      \put(22.5,45){\tiny{$(\mM_N)_{ik}$}}
    \end{picture}
    }
    }
    \caption{One loop self energies: scalar contributions. The $a$'s and $b$'s
    represent the vertex Feynman rules that can be found in
    \protect\cite{0603225,9606211}.}\label{fig:scalar self-energies} 
  \end{center}
\end{figure}
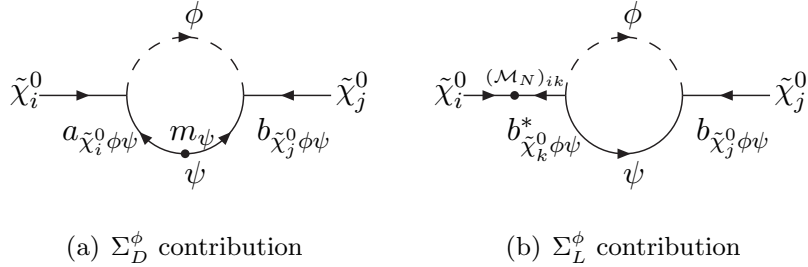

The contributions from the scalar loops in Fig.~\ref{fig:scalar self-energies} are given by
\eqa
16\pi^2(\Sigma^{\phi}_D)_{ij}(k^2) &=& -a_{\tilde{\chi}^0_i\phi\psi}b_{\tilde{\chi}^0_j\phi\psi}m_{\psi}B_0(k^2,m^2_{\psi},m^2_{\phi}),\\
16\pi^2(\Sigma^{\phi}_L)_{kj}(k^2) &=& b^*_{\tilde{\chi}^0_k\phi\psi}b_{\tilde{\chi}^0_j\phi\psi}B_1(k^2,m^2_{\psi},m^2_{\phi}).
\qea

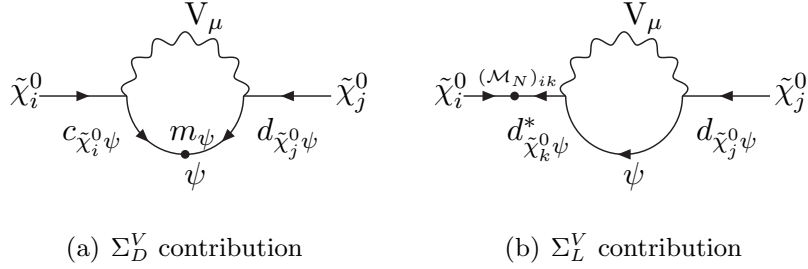
\begin{figure}[!ht]
  \begin{center}
    \scalebox{1.1}{
    \subfigure[$\Sigma^{V}_D$ contribution]{
    \begin{picture}(140,70)(0,0)
      \PhotonArc(70,40)(20,0,180){2}{6.5}
      \ArrowArc(70,40)(20,180,270)
      \ArrowArcn(70,40)(20,360,270)
      \ArrowLine(20,40)(50,40)
      \ArrowLine(120,40)(90,40)
      \Vertex(70,20){1.5}
      \put(65,25){$m_{\psi}$}
      \put(10,39){$\tilde{\chi}^0_i$}
      \put(122,39){$\tilde{\chi}^0_j$}
      \put(28,25){$c_{\tilde{\chi}^0_i\psi}$}
      \put(95,25){$d_{\tilde{\chi}^0_j\psi}$}
      \put(70,65){$\textrm{V}_{\mu}$}
      \put(70,10){$\psi$}
    \end{picture}
    }
    \subfigure[$\Sigma^{V}_L$ contribution]{
    \begin{picture}(140,70)(0,0)
      \PhotonArc(70,40)(20,0,180){2}{6.5}
      \ArrowArcn(70,40)(20,0,180)
      \ArrowLine(15,40)(32.5,40)
      \ArrowLine(50,40)(32.5,40)
      \ArrowLine(120,40)(90,40)
      \Vertex(32.5,40){1.5}
      \put(5,39){$\tilde{\chi}^0_i$}
      \put(122,39){$\tilde{\chi}^0_j$}
      \put(30,25){$d^*_{\tilde{\chi}^0_k\psi}$}
      \put(95,25){$d_{\tilde{\chi}^0_j\psi}$}
      \put(70,65){$\textrm{V}_{\mu}$}
      \put(70,10){$\psi$}
      \put(20,45){\tiny{$(\mM_N)_{ik}$}}
    \end{picture}
    }
    }
    \caption{One loop self energies: vector boson contributions.}\label{fig:vector self-energies}
  \end{center}
\end{figure}

For the vector boson contributions in Fig.~\ref{fig:vector self-energies} we
have 
\eqa
16\pi^2(\Sigma^V_D)_{ij}(k^2) &=& c_{\tilde{\chi}^0_i\psi}d_{\tilde{\chi}^0_j\psi}m_{\psi}[4B_0(k^2,m_V^2,m_{\psi}^2) \nonumber \\
&&+\Chi B_0(k^2,\Chi m_V^2,m_{\psi}^2) -B_0(k^2,m_V^2,m_{\psi}^2)],\\
16\pi^2(\Sigma^V_L)_{kj}(k^2) &=& d^*_{\tilde{\chi}^0_k\psi}d_{\tilde{\chi}^0_j\psi}[2B_1(k^2,m_{\psi}^2,m_V^2) \nonumber \\ 
&&-\Chi B_0(k^2,\Chi m_V^2,m_{\psi}^2) +B_0(k^2,m_V^2,m_{\psi}^2) \nonumber \\
&&+\frac{m^2_{\psi}-k^2}{m^2_V} (B_1(k^2,\Chi m_V^2,m_{\psi}^2)-B_1(k^2,m_V^2,m_{\psi}^2))].
\qea

In the above expressions, the PV functions are defined as in \cite{9606211}.  It should be noted that the definitions of $B_1$ in \cite{9606211} and \cite{0603225} differ by a sign.

\section{Mass insertion approximation in CPE-CPO cancellations}\label{CPECPO}
It is well known that loop contributions from CP even (CPE) and CP odd (CPO) scalars tend to cancel each other in many physical processes, because the CPO scalars couple to other fields in the same way as the CPE scalars, apart from the extra \textit{i}'s associated with the Feynman rules.  In particular, the radiative corrections to the neutrino masses involve strong cancellations among these contributions and may lead to numerical instability.  The numerical fluctuation turns out to be irrelevant for the physically interesting LNV parameter space we work in, but as we need to access regions where the light neutrino mass is typically suppressed, this issue needs to be addressed in order to obtain numerically stable results.  Another situation where this may be important is when one wishes to consider heavy neutral scalar masses.  In this case the numerical fluctuations may be comparable to the physical neutrino masses.

Instead of working directly with the cancellation, we note that in the R-parity conserving limit the cancellation between the CPE and CPO loops is exact.  Therefore it is instructive to compute the deviation from this exact cancellation arises from the LNV contributions.  We first concentrate on the CPE part.  Eq.~(\ref{eq:CPECPO blocking}) 
can be diagonalized in two stages, denoted by
\eqa
Z_R & = & \left( \begin{array}{cc}
(Z^0_H)_{2 \times 2} & 0\\
0 & (Z^0_{\tilde{\nu})_{3\times 3}} \\
\end{array} \right) \times Z^{MIA}.
\qea
In the RPC limit, $Z^{MIA}$ will be a unit matrix.  After the first stage of diagonalization, we get
\eqa
\mM^2_{CPE} &\longrightarrow& \left( \begin{array}{cc}
(\hat{M}^2_{H})_a\delta_{ab} & (Z^{0T}_{H}\sigma_H Z^0_{\tilde{\nu}})_{aj} \\
(Z^{0T}_{\tilde{\nu}}\sigma^T_H Z^0_{H})_{ib} & (\hat{M}^2_{\tilde{\nu}})_i\delta_{ij} \\
\end{array} \right),
\qea
where $\{i,j\}=1-3$, $\{a,b\}=1-2$, and the hatted quantities are diagonal
matrices.  This matrix is then diagonalized by $Z^{MIA}$, which can be
obtained by successive approximations commonly known as mass insertion
approximation (MIA).  One advantage of performing the diagonalization in these
two steps is that in the `t Hooft-Feynman gauge, the $3\times 3$ sneutrino part
of the CPE and CPO matrices are the same, despite the fact that they also
include even powers of lepton-number violating contributions.  This means that
the 
terms which contribute to the splittings between the CPE-CPO cancellations are
in $(Z^{0T}_{H}\sigma_H Z^0_{\tilde{\nu}})$.
  This procedure also allows us to
work in any flavour basis, in which the off diagonal entries
$M^2_{\tilde{\nu}}$ may be large. 

Now we can use MIA to obtain the mixing matrix $Z^{MIA}$ and the modifications to the eigenvalues.  In general, for a symmetric matrix of the form
\eqa
\mM &=& M_{\alpha}\delta_{\alpha\beta} + \epsilon_{\alpha\beta(\ne \alpha)},
\qea
where $\epsilon_{\alpha \beta} \ll M_{\alpha}$ are matrix perturbations, the mixing matrix $\mathcal{Z}$ defined by 
\eqa
\mathcal{Z}^T\mM\mathcal{Z} &=& \hat{\mM}
\qea
can be approximated to second order in $\epsilon/ M$ to be
\eqa
\begin{array}{ccc}
\mathcal{Z}_{\beta\gamma} & = & \delta_{\beta\gamma} + (\delta Z_{\epsilon})_{\beta\gamma} + (\delta Z_{\epsilon^2})_{\beta\gamma} + \mathcal{O}(\epsilon^3/M^3), \\
\end{array}
\qea
where
\eq
\begin{array}{ccl}
(\delta Z_{\epsilon})_{\beta\gamma} & = & \epsilon_{\beta\gamma}/(M_{\gamma}-M_{\beta}), \\
(\delta Z_{\epsilon^2})_{\beta\gamma} & = & -\sum_{\rho}\epsilon_{\beta\rho}\epsilon_{\rho\gamma}/2(M_{\beta}-M_{\rho})(M_{\gamma}-M_{\rho})\delta_{\beta\gamma} \\
&& +\sum_{\rho}\epsilon_{\beta\rho}\epsilon_{\rho\gamma(\ne \beta)}/2(M_{\gamma}-M_{\beta})(M_{\gamma}-M_{\rho}),
\end{array}
\qe
and $\hat{\mM}$ is a diagonal matrix.  From here it is easy to obtain an
approximate correction to the eigenvalues
\eq
\delta M_{\alpha} = \hat{\mM}_{\alpha} - M_{\alpha} = \sum_{\rho}\frac{\epsilon_{\alpha\rho}\epsilon_{\rho\alpha}}{M_{\alpha}-M_{\rho}}.
\qe

This can be used to obtain the analytic expansion of Eq.~(3.5) in \cite{0603225}.  In the basis in which we perform our diagonalization of $\mM'_N$, this is given by
\eqa
(\Sigma^{\nu\nu}_D)_{ij} &=& -\sum_{s=1}^{5}\sum_{r=1}^{7}\frac{m_{\kappa^0_r}}{(4\pi)^2}\Big(\frac{e}{2c_W}Z_{N1r}-\frac{e}{2c_W}Z_{N1r}\Big)^2\nonumber\\
&&\Big[Z_{R(2+i)s}Z_{R(2+j)s}B_0(0,m^2_{H^0_s},m^2_{\kappa^0_r})-Z_{A(2+i)s}Z_{A(2+j)s}B_0(0,m^2_{A^0_s},m^2_{\kappa^0_r})\Big] \label{eqtwo}
\qea
where $m_{H^0_{1,\ldots,5}}$ and $m_{A^0_{1,\ldots,5}}$ are the CPE and CPO neutral
scalars, $m_{\kappa^0_{1,\ldots,7}}$ are the neutral fermions propagating in the
loop, and  $Z_R$ and $Z_A$ are the CPE and CPO scalar mixing matrices
respectively.  Specifically, we seek an expansion of the expression in square
brackets in Eq.~(\ref{eqtwo})
\eq
\begin{array}{rl}
&Z_{R(2+i)s}Z_{R(2+j)s}B_0(0,m^2_{H^0_s},m^2_{\kappa^0_r})-Z_{A(2+i)s}Z_{A(2+j)s}B_0(0,m^2_{A^0_s},m^2_{\kappa^0_r})
 \\
 \approx & Z_{R(2+i)h}Z_{R(2+j)h}B_0(0,m^2_{h^0},m^2_{\kappa^0_r})+Z_{R(2+i)H}Z_{R(2+j)H}B_0(0,m^2_{H^0},m^2_{\kappa^0_r}) \\
&- Z_{A(2+i)G}Z_{A(2+j)G}B_0(0,m^2_{G^0},m^2_{\kappa^0_r})-Z_{A(2+i)A}Z_{A(2+j)A}B_0(0,m^2_{A^0},m^2_{\kappa^0_r}) \\
&+ (Z_{R(2+i)\tilde{\nu}_s}Z_{R(2+j)\tilde{\nu}_s}-Z_{A(2+i)\tilde{\nu}_s}Z_{A(2+j)\tilde{\nu}_s})B_0(0,m^2_{\tilde{\nu}^{CPO}_s},m^2_{\kappa^0_r}) \\
&+ (m^2_{\tilde{\nu}^{CPE}_s} - m^2_{\tilde{\nu}^{CPO}_s})Z_{R(2+i)\tilde{\nu}_s}Z_{R(2+j)\tilde{\nu}_s}C_0(m^2_{\tilde{\nu}^{CPO}_s},m^2_{\tilde{\nu}^{CPO}_s},m^2_{\kappa^0_r})\\
&+ \textrm{higher order terms.}
\end{array}
\qe
One can then substitute in the results from the MIA to obtain an analytic
approximation to $\mathcal{O}(\sigma^2/M^2)$.  It is rather lengthy to show
the full analytic expansion here.  We 
merely  note that in the limit of zero
sneutrino vevs and no sneutrino mixing, the above expression agrees with
Eq.~(4.10) in \cite{0603225}, apart from the off-diagonal contributions of
$\delta Z_{\epsilon^2}$, which is not included in Ref.~\cite{0603225} but is
important 
for our purpose. 

\section{Calculation of $l^I \to l^J \gamma$}\label{llgamma}
In this section we give an explicit calculation of the branching ratio of leptonic FCNC: $BR(l^I \to l^J\gamma)$.  For the phase space calculation, 4 component spinor convention as in \cite{0707.3718} is used.  For the calculation of the loop integrals, 2 component spinor notation is used instead.  See the Appendix of \cite{0603225} and references therein for details on the 2 spinor notation calculations.

The effective Lagrangian for decay $l^I \to l^J \gamma$ is given by
\eq
\mathcal{L}_{eff}=e\bar{l}^J\sigma^{\mu\nu}(\Lambda_L P_L + \Lambda_R P_R)l^I F_{\mu\nu},
\qe
where $\sigma^{\mu\nu}=\frac{i}{2}[\gamma^{\mu},\gamma^{\nu}]$, and $P_{L/R}=\frac{1}{2}(1\mp \gamma_5)$.  $\Lambda_{L(R)}$ are obtained from finite part of one loop diagrams to be described later.  In this notation, the matrix element for the process is given by
\eqa
M=2e\bar{u}_{l^J}\sigma^{\mu\nu}(\Lambda_L P_L + \Lambda_R P_R)u_{l^I}\epsilon_{\mu}q_{\nu},
\qea
where $u_{l^I}, {\bar u}_{l^J}$ are 4-component Dirac spinors.
Summing over final state spins in the matrix element squared, and averaging
over the initial $l^I$ spin, we obtain 
\eqa
K&=&\frac{1}{2}\sum_{\epsilon,s_{l^I},s_{l^J}}|M|^2\nonumber\\
&=& 2e^2Tr\{\sigma^{\mu\nu}(\Lambda_L P_L + \Lambda_R P_R)(\not\!\!p_{l^I}+m_{l^I})(\Lambda_L^* P_R + \Lambda_R^* P_L)\nonumber\\
&&\quad\qquad\sigma^{\rho\sigma}(\not\!\!p_{l^J}+m_{l^J})(-\eta_{\mu\rho})q_{\nu}q_{\sigma}\}
\nonumber \\
&=&16e^2\Big(|\Lambda_L|^2 + |\Lambda_R|^2\Big)(q\cdot p_{l^I})(q\cdot p_{l^J}).
\qea
The decay width of $l^I \to l^J \gamma$ is given by the standard formula
\eqa
\Gamma(l^I \to l^J \gamma) &=& \frac{1}{2 m_{l^I}}\int\frac{d^3 p_{l^J}}{(2\pi)^3 2E_{l^J}}\frac{d^3q}{(2\pi)^3 2E_q}(2\pi)^4\delta(p_{l^I}-p_{l^J}-q)K \nonumber \\
&=&\frac{e^2}{4\pi}m_{l^I}^3
(|\Lambda_L|^2 + |\Lambda_R|^2),
\qea
Using the result $\Gamma(l^I \to l^J \bar{\nu}^{J}\nu^I)=\textrm{G}_F^2m_I^5/192\pi^3$ 
(see for example Ref.~\cite{ChengAndLi}), the branching ratio is then related to 
$BR(l^I \to l^J \bar{\nu}^{J}\nu^I)$ by
\eq
BR(l^I \to l^J \gamma)=\frac{48\pi^2e^2}{m_{l^I}^2G_F^2}
(|\Lambda_L|^2 + |\Lambda_R|^2) BR(l^I \to l^J \bar{\nu}^{J} \nu^I),
\qe
where the branching ratios $BR(l^I \to l^J \bar{\nu}^{J} \nu^I)$ are given by \cite{PDG}
\eqa
BR(\tau \to \mu \bar{\nu}^{\mu} \nu^{\tau})&=& 0.1736\pm 0.0005 \nonumber \\
BR(\tau \to e \bar{\nu}^{e} \nu^{\tau})&=& 0.1784\pm 0.0005 \nonumber \\
BR(\mu \to e \bar{\nu}^{e} \nu^{\mu})&\simeq& 1.
\qea

The calculation of the Wilson coefficients $\Lambda_L$ and $\Lambda_R$ can be
performed in the 4- or the 2-component formalisms.  To compute the loop integrals it probably is slightly easier to use 4 component conventions.  On the other hand, the 2 component formalism is more transparent, especially to visualize the helicity flips from the fermion mass insertions.  We will use the latter method.

\FIGURE{
  \scalebox{0.9}{
    \begin{picture}(140,100)(0,0)
      \DashArrowArcn(70,40)(20,180,360){5}
      \ArrowArcn(70,40)(20,270,180)
      \ArrowArc(70,40)(20,270,360)
      \ArrowLine(20,40)(50,40)
      \ArrowLine(120,40)(90,40)
      \Photon(90,60)(115,75){3}{5}
      \Vertex(70,20){1.5}
      \put(65,27.5){$m_{\psi}$}
      \put(10,35){$l^I$}
      \put(125,35){$l^{cJ}$}
      \put(30,25){$a_{I\phi\psi}$}
      \put(95,25){$b_{J\phi\psi}$}
      \put(70,65){$\phi$}
      \put(70,10){$\psi$}
    \end{picture}
    \begin{picture}(140,100)(0,0)
      \DashArrowArcn(70,40)(20,180,360){5}
      \ArrowArcn(70,40)(20,0,180)
      \ArrowLine(20,40)(50,40)
      \ArrowLine(90,40)(105,40)
      \ArrowLine(120,40)(105,40)
      \Photon(90,60)(115,75){3}{5}
      \Vertex(105,40){1.5}
      \put(102.5,45){$m_J$}
      \put(10,35){$l^I$}
      \put(125,35){$l^{cJ}$}
      \put(30,25){$a_{I\phi\psi}$}
      \put(95,25){$a^*_{J\phi\psi}$}
      \put(70,65){$\phi$}
      \put(70,10){$\psi$}
    \end{picture}
    \begin{picture}(140,100)(0,0)
      \DashArrowArcn(70,40)(20,180,360){5}
      \ArrowArc(70,40)(20,180,0)
      \ArrowLine(20,40)(35,40)
      \ArrowLine(50,40)(35,40)
      \ArrowLine(120,40)(90,40)
      \Photon(90,60)(115,75){3}{5}
      \Vertex(35,40){1.5}
      \put(32.5,45){$m_I$}
      \put(10,35){$l^I$}
      \put(125,35){$l^{cJ}$}
      \put(30,25){$b^*_{I\phi\psi}$}
      \put(95,25){$b_{J\phi\psi}$}
      \put(70,65){$\phi$}
      \put(70,10){$\psi$}
    \end{picture}
  }
  \caption{$l^I \to l^J\gamma$ decay} \label{fig:llgamma diagram}
}

The loop integrals for $\Lambda_L$ are displayed in Fig.~\ref{fig:llgamma
  diagram}.  For a loop with scalar $\phi$ and fermion $\psi$, the
$a_{I\phi\psi}$ couplings correspond to a left-handed, negatively charged lepton $l^I$ from the SU(2) doublet which flows into the vertex, whereas the $b_{I\phi\psi}$ couplings correspond to \textit{left}-handed, positively charged \textit{anti}-lepton $l^{cI}$ from the SU(2) singlet, again flowing into the vertex.  The $\Lambda_R$ diagrams are similar to the $\Lambda_L$ ones, and can be obtained by interchanging $a$ with $b^*$ and by reversing the helicity flow.  Further details can be found in \cite{0603225} and \cite{0610406}.

The analytic expressions for $l^I \to l^J \gamma$ are given by:
\eqa
\Lambda_L^{IJ}&=&\frac{1}{2}\frac{n_c}{(4\pi)^2}
\Big\{a_{I\phi \psi}b_{J\phi \psi}m_{\psi}
\Big[\frac{Q_{\phi}}{m^2_{\phi}}F_4\Big(\frac{m^2_{\psi}}{m^2_{\phi}}\Big)
  -\frac{Q_{\psi}}{m^2_{\phi}}F_3\Big(\frac{m^2_{\psi}}{m^2_{\phi}}\Big)\Big] \nonumber \\
&&+(a_{I\phi \psi}a^*_{J\phi \psi}m_{J}+b^*_{I\phi \psi}b_{J\phi \psi}m_{I})
\Big[\frac{Q_{\phi}}{m^2_{\phi}}F_2\Big(\frac{m^2_{\psi}}{m^2_{\phi}}\Big)
-\frac{Q_{\psi}}{m^2_{\phi}}F_1\Big(\frac{m^2_{\psi}}{m^2_{\phi}}\Big)\Big] \Big\}\qquad \\
\Lambda_R^{IJ}&=&\frac{1}{2}\frac{n_c}{(4\pi)^2}
\Big\{b^*_{I\phi \psi}a^*_{J\phi \psi}m_{\psi}
\Big[\frac{Q_{\phi}}{m^2_{\phi}}F_4\Big(\frac{m^2_{\psi}}{m^2_{\phi}}\Big)
  -\frac{Q_{\psi}}{m^2_{\phi}}F_3\Big(\frac{m^2_{\psi}}{m^2_{\phi}}\Big)\Big] \nonumber \\
&&+(b^*_{I\phi \psi}b_{J\phi \psi}m_{J}+a_{I\phi \psi}a^*_{J\phi \psi}m_{I})
\Big[\frac{Q_{\phi}}{m^2_{\phi}}F_2\Big(\frac{m^2_{\psi}}{m^2_{\phi}}\Big)
-\frac{Q_{\psi}}{m^2_{\phi}}F_1\Big(\frac{m^2_{\psi}}{m^2_{\phi}}\Big)\Big] \Big\},\qquad
\qea
where $Q_{\psi}$ and $Q_{\phi}$ are the electric charges of the scalar and fermion respectively, flowing away from the first vertex in the decay.  $n_c$ is the number of colour in the loop.  
The loop integrals are given by:
\eqa\label{eq: loopintegral}
F_1(x)&=&\frac{1}{12(1-x)^4}[2+3x-6x^2+x^3+6x\textrm{ln}x],\nonumber\\
F_2(x)&=&\frac{1}{12(1-x)^4}[1-6x+3x^2+2x^3-6x^2\textrm{ln}x],\nonumber\\
F_3(x)&=&\frac{1}{2(1-x)^3}[-3+4x-x^2-2\textrm{ln}x],\nonumber\\
F_4(x)&=&\frac{1}{2(1-x)^3}[1-x^2+2x\textrm{ln}x].
\qea

Strictly speaking, the above loop integrals are only valid if the external fermion masses are negligible.  Contributions from LNV terms typically involve SM fermions in the loop, so the external fermion masses cannot be neglected for a full calculation.  However, because the loop scalar remains massive, it is possible to obtain corrections in powers of $m^2_l/m^2_{\phi}$, which are negligible for our purpose\footnote{In fact all the $x$'s in Eq.~(\ref{eq: loopintegral}) can be neglected for the LNV loops, apart from a possible top quark in the loop.  We include them only for the purpose of comparing with the literature in the R-parity conserving limit, when both the scalar and the fermion in the loop are heavy and so the $x$'s cannot be neglected.}.  To be more explicit, we use the integral $F_3$ as an example.  The full integral is given by
\eqa
\frac{1}{m^2_{\phi}}F_3 &=& \int_0^1dz \int_0^{1-z}dy\frac{1-z}{z(m^2_{\phi}-ym^2_{l^I}-(1-y-z)m^2_{l^J})+(1-z)m^2_{\psi}}.
\qea
Thus the approximation in Eq.~(\ref{eq: loopintegral}) is valid up to $\sim{\mathcal O}(m^2_l/m^2_{\phi})$.  Readers who are interested in applications of the full loop integrals, for example in a supersymmetric theory where the loop scalar mass can be light, may consult \cite{9908443} for instance.

We have checked explicitly that the above expressions are in agreement with \cite{0707.3718} and \cite{0004067} (in the context of $b\to s\gamma$)\footnote{The sign of the log term in $F_4(x)$ as well as the overall signs of the terms involving $F_1(x)$ and $F_2(x)$ are different compared to \cite{0610406}.}.  We also checked that the expression for $(g-2)_{\mu}$ as well as the numerical results of the SPS benchmarks in the RPC limit agree with \cite{0609168}.

\end{document}